\documentclass[twocolumn,showpacs,amsmath,amssymb]{revtex4}
\input{epsf}

\usepackage{graphicx}
\usepackage{array}
\usepackage{longtable}
\usepackage{rotating,booktabs}
\usepackage{booktabs,threeparttable}
\usepackage{bm}
\usepackage{url}

\usepackage{color}

\begin{document}

\title{Relativistic calculations of quasi-one-electron atoms and ions \\
       using Laguerre and Slater spinors}

\author{Jun Jiang}
\email{phyjiang@yeah.net}
\affiliation{Key Laboratory of Atomic and Molecular Physics and Functional Materials of Gansu Province, College of Physics and Electronic Engineering, Northwest Normal University, Lanzhou 730070, P. R. China}
\affiliation{School of Engineering, Charles Darwin University, Darwin NT 0909, Australia}

\author{J. Mitroy}
\altaffiliation[Deceased]{.}
\affiliation{School of Engineering, Charles Darwin University, Darwin NT 0909, Australia}

\author{Yongjun Cheng}
\affiliation{Academy of Fundamental and Interdisciplinary Science, Harbin Institute of Technology, Harbin 150080, P. R. China}
\affiliation{School of Engineering, Charles Darwin University, Darwin NT 0909, Australia}

\author{Michael W. J. Bromley}
\homepage{https://www.smp.uq.edu.au/people/brom/}
\affiliation{School of Mathematics and Physics, The University of Queensland, Brisbane, Queensland 4075, Australia}

\date{\today}

\begin{abstract}
A relativistic description of the structure of heavy alkali atoms and
alkali-like ions using S-spinors and L-spinors has been developed.
The core wavefunction is defined by a Dirac-Fock calculation using an S-spinors
basis.  The S-spinor basis is then supplemented by a large set of L-spinors for
the calculation of the valence wavefunction in a frozen-core model.
The numerical stability of the L-spinor approach is demonstrated by computing
the energies and decay rates of several low-lying hydrogen eigenstates, along
with the polarizabilities of a $Z=60$ hydrogenic ion.
The approach is then applied to calculate the dynamic polarizabilities
of the $5s$, $4d$ and $5p$ states of Sr$^+$.  The magic wavelengths
at which the Stark shifts between different pairs of transitions are zero
are computed.  Determination of the magic wavelengths for the
$5s \to 4d_{\frac32}$ and $5s \to 4d_{\frac52}$ transitions near $417$~nm
(near the wavelength for the $5s \to 5p_j$ transitions) would allow
a determination of the oscillator strength ratio for
the $5s \to 5p_{\frac12}$ and $5s \to 5p_{\frac32}$ transitions.

\end{abstract}

 \maketitle

\section{Introduction}

This paper describes the development and application of a
relativistic model for atomic structure.  The basic strategy of the
model is to partition the atom into valence and core electrons.
The core electrons will be represented by orbitals obtained from
Dirac-Fock calculations.  The wave function for the valence electrons
will be computed by expanding the wave function as a linear combination
of Laguerre function spinors (L-spinors) and Slater function spinors
(S-spinors)~\cite{quiney89a,grant00a,grant07a}.  The direct and exchange interactions
between the core and valence electrons can be computed without approximation.
Core-valence correlations can be represented by simply introducing semi-empirical
core polarization potentials which are tuned to ensure that the energies for the
valence electrons agree with experiment~\cite{mitroy88d,mitroy93d,mitroy03f}.

The motivation for this methodology is based on the success of similar
methodologies in computing atomic properties of light atoms, namely
non-relativistic configuration interaction with a
semi-empirical core potential method
(CICP)~\cite{migdalek78a,mitroy88d,mitroy93d,mitroy03f}.  As a
recent example, the dipole polarizability of the Si$^{2+}$ ion computed
with a similar methodology is $11.688$~$a_0^3$~\cite{mitroy08k}.  An
analysis of a resonant excitation stark ionization spectroscopy (RESIS)~\cite{lundeen05a}
experiment give $11.669(9)$~$a_0^3$~\cite{komara05a,mitroy08k}
while a very sophisticated relativistic configuration interaction with many body
perturbation theory calculation (MBPT) gave $11.670(13)$~$a_0^3$~\cite{safronova12e}.
Numerous other examples of very good agreement of the semi-empirical
method with the most advanced \textit{ab-initio} theoretical models for oscillator
strengths and polarizabilities can be found in Ref.~\cite{mitroy09a,safronova11c,porsev12a}.

There are a number of reasons for the success of the relativistic semi-empirical approach.
Firstly, this approach is based on the \textit{ab-initio} Dirac-Fock (DF) calculation to define
the core.  Secondly, tuning energies to experimental values leads to
wave functions that have the correct asymptotic decay at long distances
from the nucleus.  The multipole matrix elements needed for oscillator strength
and polarizability calculations tend to be dominated by the large-$r$ form of
the wave function. Finally, partitioning the wave function into
frozen-core electrons and an active valence electron reduces the
equation for the wave function and energies into one equation
that admits a close to exact numerical solution, here using a
large (orthogonal) Laguerre basis.

It should be noted that the DF+core-polarization method adopted here has
been extensively used by Migdalek and co-workers to calculate the oscillator
strengths of many atoms~\cite{migdalek84a,migdalek86a,migdalek93a,migdalek07a}.
They solved the radial equations numerically~\cite{migdalek76a}, and they
typically restricted their transitions to between those of the low-lying states.
Here, we employ basis sets which enables the calculation of
transition matrix elements between both the bound states and the
continuum (pseudostates).  This enables us here to compute atomic
polarizabilities~\cite{mitroy10a}, where the continuum makes a
significant contribution~\cite{mitroy03f}.

The present work gives a brief description of the strategy adopted
to convert an existing non-relativistic Hartree-Fock (HF)
program~\cite{mitroy99f} into a relativistic DF program.
Next, the technical details for performing calculations for one valence
electron atoms and ions are discussed. These methods are then applied
to the solution of hydrogen and hydrogenic atoms as a test for evaluation.

The main results presented are the oscillator strengths, and static and dynamic
polarizabilities for the low-lying states of Sr$^+$ ions.  In addition
some of the magic wavelengths for $5s-5p_J$ and $5s-4d_J$ transitions are
presented, at which the ac-Stark shift of the transition energy is zero.
The static polarizabilities of Sr$^+$ can be used to
estimate frequency shifts of $5s-4d_J$ clock transitions due to background fields
such as blackbody radiation shifts~\cite{dube14a}. The magic wavelengths can
be used, for example, for high-precision trapping measurements~\cite{madej12a,ludlow15a}

\section{Formulation and Validations}

The single-electron Dirac equation can be written as,
\begin{eqnarray}
H\Psi(\textbf{r})=E\Psi(\textbf{r})\,,
\label{eqn:dirac}
\end{eqnarray}
where the Hamiltonian
\begin{eqnarray}
H=c\bm{\alpha}\cdot\bm{p}+\beta c^{2}-\frac{Z}{r} + V_{core} \,,
\label{eqn:diracham}
\end{eqnarray}
$\bm{p}$ is the momentum operator,
$\bm{\alpha}$ and $\beta$ are $4\times 4$
matrices of the Dirac operators~\cite{kaneko77a}.
The $V_{core}$ represents the valence electron-core electrons
interaction, and is described shortly.

We have two separate codes that we present the first results from here.
The first is the DF calculation, which generates the closed-shell
orbitals using purely Slater-type orbitals.  The second code
solves for a single valence electron orbiting the closed-shell
using a mixture of the Slater-type orbitals produced by the
first code with additional Laguerre-type orbitals to describe the
valence electronic structure and continuum physics.

\subsection{Calculations of core orbitals}

The starting point of a calculation involving closed shells
is the DF calculation for the core state of the atoms.
The DF equations are closely related to the HF equations.
The atomic Schr{\"o}dinger Hamiltonian is replaced by the Dirac-Coulomb
Hamiltonian and the single particle orbitals are now 4-component spinors
with a large and a small component.

The strategy used to generate a DF wave function is to adapt
an existing HF program~\cite{mitroy99f}
which expands the orbitals as a linear combination of
Slater (or Gaussian) type orbitals.  The first stage of the modification
is to generate the angular representation of the orbitals from
$\ell \to \ell,j$ representation.

The next stage is to write each orbital in terms of $S$-spinors.
Each orbital wavefunction can be written as
\begin{eqnarray}
\psi_{n \kappa m}(\textbf{r})=\frac{1}{r} \left (
\begin{array}{r}
P_{n \kappa}(r)\Omega_{\kappa m}(\hat{\textbf{r}})\\
iQ_{n \kappa}(r)\Omega_{-\kappa m}(\hat{\textbf{r}})\\
\end{array}
\right ) \,,
\label{eqn:orbitals}
\end{eqnarray}
where $\kappa$ is the relativistic angular quantum number
which is connected to  the total angular momentum quantum number $j$
and the orbital angular momentum quantum number $\ell$,
\begin{equation}
\kappa=\ell(\ell+1)-j(j+1)-\frac14 \,.
\label{eqn:kappa}
\end{equation}
$P_{n\kappa}(r)$ and $Q_{n\kappa}(r)$ represent the large and
small components of radial wavefunction, and
$\Omega_{\kappa m}(\hat{\textbf{r}})$ and
$\Omega_{-\kappa m}(\hat{\textbf{r}})$ are
the angular components.

The radial wavefunctions $P_{n\kappa}(r)$ and $Q_{n\kappa}(r)$ are
expanded as $N$-terms in an S-spinor basis
\begin{equation}
P_{n\kappa}(r)=\sum_{i=1}^{N} p_i \phi_{i,\kappa}^{P}(r) , \quad
Q_{n\kappa}(r)=\sum_{i=1}^{N} q_i \phi_{i,\kappa}^{Q}(r) ,
\end{equation}
where the superscript $P$ and $Q$ identify the ``large" and ``small"
components of the Dirac spinor in a conventional way.

Although it is common to formally sub-divide
the basis functions into small and large type functions
and explicitly recognize this when casting the DF equations
into operational form~\cite{grant07a},
that approach is not adopted in the present paper.
Instead, each orbital has a label identifying it as
being of a large or small component in the present code.
These labels are taken into account
when computing the matrix elements of the DF Hamiltonian.  This
approach is adopted since minimal modifications are needed
for those parts of the program that construct and diagonalize
the Hamiltonian.  In effect, information about the
spinor construction is confined to those parts of the program
that evaluate the matrix elements of the basis functions.

S-spinors are generalizations of Slater type orbitals (STO) adapted to
relativistic systems.  The first modification is the inclusion of
a radial $r^\gamma$ pre-factor with
\begin{eqnarray}
\gamma(\kappa) = \sqrt{\kappa^2-Z^2/c^2} \,
\end{eqnarray}
to ensure these functions have the correct asymptotic form at origin.
Here, $Z$ is the atomic number and we adopt $c = 137.0359991$ as the
speed of light (in atomic units).

The second modification includes choosing the large and small component basis
functions to approximately satisfy the kinetic balance condition~\cite{grant00a}.
The unnormalized radial components are written as,
\begin{eqnarray}
\phi_{i,\kappa}^{P,Q}(r)= r^\gamma e^{-\lambda_i r}
\end{eqnarray}
for orbitals with $\kappa < 0$, and
\begin{eqnarray}
\phi_{i,\kappa}^{P,Q}(r)= A_{P,Q}
r^\gamma e^{-\lambda_i r}+ \lambda r^{\gamma+1} e^{-\lambda_i r}
\end{eqnarray}
for orbitals with $\kappa>0$, where
\begin{equation}
A_P=\frac{\left(\kappa+1-\sqrt{\kappa^2+2\gamma+1}\right)\left(2\gamma+1\right)}{2\left(\sqrt{\kappa^2+2\gamma+1}-\kappa\right)}
\end{equation}
for the large components and
\begin{equation}
A_Q=\frac{\left(\kappa-1-\sqrt{\kappa^2+2\gamma+1}\right)\left(2\gamma+1\right)}{2\left(\sqrt{\kappa^2+2\gamma+1}-\kappa\right)}
\end{equation}
for the small components.

\subsubsection{Numerical test: energy of closed-shell atoms}

A DF basis set is formed as a collection of S-spinors
with positive real exponents
$\{\lambda_i\}$
and coefficients $\{p_i\}$ and $\{q_i\}$
$\forall i=1,2,\hdots,N_S$.
that undergo variational optimization.
The S-spinor for the orbitals with $\kappa <0$ has a very
simple form. The radial prefactor did not allow for additional powers
of $r$ as prefactors. This is distinct from the related STO basis
sets used for non-relativistic calculations which
usually have radial prefactors with a variety of powers of
$r$~\cite{clementi74a}. In our calculations,
the S-spinor basis sets used are based on non-relativistic
basis sets. An STO basis with all functions restricted to
$n = \ell + 1$ was optimized for the non-relativistic calculation.
Once the optimization was complete, this was modified by the
replacement $n$ $\to$ $\sqrt{\kappa^2 - Z^2/c^2}$ for S-spinors.
This is based on the form of the exact wave functions for
$\kappa < 0$.
No further minor optimizations is undertaken as the
relativistic scf calculations are time comsuming to do.

Table~\ref{tab:dfenergies} gives DF energies computed using S-spinor
basis and numerical DF energies computed using GRASP92~\cite{parpia96a}.
It can be seen that the two sets of energies are
in agreement with each other to at least six significant digits.
Note that GRASP92 uses a finite difference method, so such
differences are expected.
See Supplemental Table~I, Table~II, and Table~III for lists of
the basis exponents.
\begin{table}
\caption{\label{tab:dfenergies} Comparison of numerical DF energies (in a.u.)
of several closed-shell atoms and ions as computed with various S-spinor basis sets
using the present S-spinor program and the GRASP92 program~\cite{parpia96a}.
The notation $a[b]$ indicates $a \times 10^b$.
The underlines denote the digits which are different from the
two programs.}
\begin{ruledtabular}
\begin{tabular}{llll}
Atom/Ion & Basis Set         &  S-spinor                    &   GRASP92            \\
\hline
Li$^+$   & $7s$              & $-$7.237205\underline{25}    & $-$7.23720552    \\
Na$^+$   & $7s$,$4p$         & $-$1.61895\underline{877}[2] & $-$1.61895968[2] \\
K$^+$    & $8s$,$7p$         & $-$6.01378\underline{956}[2] & $-$6.01379058[2] \\
Rb$^+$   & $11s$,$8p$,$5d$   & $-$2.9796932\underline{3}[3] & $-$2.97969324[3] \\
Cs$^+$   & $14s$,$12p$,$10d$ & $-$7.78694\underline{367}[3] & $-$7.78694284[3] \\
Ne       & $8s$,$5p$         & $-$1.28691\underline{836}[2] & $-$1.28691970[2] \\
Ar       & $11s$,$9p$        & $-$5.286844\underline{41}[2] & $-$5.28684451[2] \\
Kr       & $10s$,$9p$,$5d$   & $-$2.78888\underline{845}[3] & $-$2.78888486[3] \\
Xe       & $14s$,$12p$,$9d$  & $-$7.447162\underline{55}[3] & $-$7.44716272[3] \\
Be$^{2+}$& $5s$              & $-$1.361399\underline{56}[1] & $-$1.36140014[1] \\
Mg$^{2+}$& $9s$,$6p$         & $-$1.991501\underline{19}[2] & $-$1.99150137[2] \\
Ca$^{2+}$& $10s$,$9p$        & $-$6.791050\underline{26}[2] & $-$6.79105063[2] \\
Sr$^{2+}$& $12s$,$10p$,$5d$  & $-$3.17755\underline{410}[3] & $-$3.17755362[3] \\
Ba$^{2+}$& $15s$,$14p$,$10d$ & $-$8.13548\underline{402}[3] & $-$8.13548296[3] \\
\end{tabular}
\end{ruledtabular}
\end{table}

\subsection{Calculation of valence orbitals}

The orbitals for the valence electrons are written as
linear combinations of S-spinors and L-spinors.
L-spinors are generalizations of Laguerre type orbitals~\cite{bromley01a} adapted to
relativistic systems, and they are derived from the relativistic analogues of
Coulomb Sturmians~\cite{grant00a}. The (unnormalized) L-spinors are written as
 \begin{eqnarray}
 \phi_{i,\kappa}^P (r) = r^\gamma e^{-\lambda_i r}
  \Big \{ &(\delta_{n_i,0}-1) L_{n_i-1}^{2\gamma}(2\lambda_i r)  \nonumber \\
         &+ B L_{n_i}^{2\gamma}(2\lambda_i r)\Big \}
\end{eqnarray}
and
\begin{eqnarray}
\phi_{i,\kappa}^Q (r)= r^\gamma e^{-\lambda_i r}
\Big \{ &(\delta_{n_i,0}-1) L_{n_i-1}^{2\gamma}(2\lambda_i r) \nonumber \\
&-BL_{n_i}^{2\gamma}(2\lambda_i r)\Big \} \, ,
\end{eqnarray}
where the balanced coefficient
\begin{equation}
B=\frac{\sqrt{n_i^2+2n_i\gamma+\kappa^2} - \kappa}{n_i+2\gamma} \, ,
\end{equation}
with $n_i$ being a non-negative integer
($n_i \geqslant 0 $ for $\kappa < 0 $ and $n_i \geqslant 1 $ for $\kappa > 0 $),
The $L_n^{\alpha}$ are Laguerre polynomials~\cite{abramowitz72a}
which are computed using the recursion relation
\begin{equation}
     L^{\alpha}_{n+1}(x) = \frac{(2n+\alpha+1-x)}{(n+1)}L^{\alpha}_n(x)
                        - \frac{(n+\alpha)}{(n+1)} L^{\alpha}_{n-1}(x),
\end{equation}
with $L^{\alpha}_0(x) = 1$ and $L^{\alpha}_1(x) = 1+\alpha - x$.
In our single-valence electron calculations, we always choose $2N$ L-spinor
orbitals which include $N$ large component orbitals and $N$ small component orbitals.

The radial Dirac equation, Eqn.~\ref{eqn:dirac},
can be solved as a (real, symmetric) matrix eigenproblem,
with the resulting set of $N$ eigenfunctions
\begin{equation}
   \Psi_{I}({\bf r}) = \sum_{n_I=1}^{N} c_{n_I} \psi_{n_I \kappa_I m_I}(\textbf{r}) \, ,
\end{equation}
where $I \in 1, \hdots, N$.
In order to compare with non-relativistic calculations, we
replace the energy $E$ by $\varepsilon=E-mc^{2}$,
where $m$ is the mass of the electron ($m=1$ in atomic units).

\subsubsection{Numerical test: energy of hydrogen atom}

Our code was first tested by diagonalizing the
ground state of hydrogen (ie. $Z = 1$, $V_{core} = 0$)
with $N=50$ L-spinors.
A value of $\lambda = 2.0$ was chosen for the $s$ orbitals
and $\lambda=1.0$ for other orbitals
($\lambda = 1.0$ would correspond to the exact hydrogen ground state).
The results for several eigenstates of
hydrogen are shown in Table~\ref{tab:hydrogengamma} and compare
well with the NIST (experimental) values~\cite{nistasd500},
given that we are using the infinite proton mass approximation.
These calculations were performed in quadruple precision arithmetic
(also shown are their decay rates: these are discussed in the next section).
\begin{table}
\caption{\label{tab:hydrogengamma} Theoretical (RCI) energies ($\varepsilon$ in Hartree)
and separated E1 ($\Gamma^{(1)}$) and E2 ($\Gamma^{(2)}$) decay rates (in 1/nanoseconds) for
several eigenstates of hydrogen.
The underlines denote the digits which are different from the
NIST tabulation~\cite{nistasd500}.
The notation $a[b]$ indicates $a \times 10^b$.
}
\begin{ruledtabular}
\begin {tabular}{ccccc}
      $I$ &  $j_I$  & $\varepsilon_I$ & $\Gamma_I^{(1)}$ & $\Gamma_I^{(2)}$ \\
\hline
$1s$ & $\frac12$ &  $-0.500\underline{0066566}$   & $--$  & $--$  \\
$2s$ & $\frac12$ &  $-0.1250\underline{020802}$   & $--$  & $--$  \\
$2p$ & $\frac12$ &  $-0.1250\underline{020802}$   & $6.26\underline{831}[8]$  & $--$  \\
     & $\frac32$ &  $-0.1250\underline{004160}$   & $6.26\underline{838}[8]$  & $1.310[-22]$  \\
$3s$ & $\frac12$ &  $-0.0555\underline{5629518}$  & $6.31\underline{771}[6]$  & $--$  \\
$3p$ & $\frac12$ &  $-0.0555\underline{5629516}$  & $1.89\underline{801}[8]$  & $23.9212$  \\
     & $\frac32$ &  $-0.0555\underline{5580208}$  & $1.89\underline{807}[8]$  & $23.9214$  \\
$3d$ & $\frac32$ &  $-0.0555\underline{5580210}$  & $6.46\underline{874}[7]$  & $645.117$  \\
     & $\frac52$ &  $-0.0555\underline{5563772}$  & $6.46\underline{864}[7]$  & $645.125$  \\
$4s$ & $\frac12$ &  $-0.0312\underline{5033803}$  & $4.41\underline{642}[6]$  & $1.02876$  \\
$4p$ & $\frac12$ &  $-0.0312\underline{5033803}$  & $8.13\underline{100}[7]$  & $12.8530$  \\
     & $\frac32$ &  $-0.0312\underline{5013001}$  & $8.13\underline{129}[7]$  & $12.8534$  \\
$4d$ & $\frac32$ &  $-0.0312\underline{5013003}$  & $2.76\underline{784}[7]$  & $337.072$  \\
     & $\frac52$ &  $-0.0312\underline{5006066}$  & $2.76\underline{779}[7]$  & $337.078$  \\
$4f$ & $\frac72$ &  $-0.0312\underline{5006067}$  & $1.37\underline{955}[7]$  & $67.6017$  \\
     & $\frac52$ &  $-0.0312\underline{5002601}$  & $1.37\underline{954}[7]$  & $67.6014$  \\
\end{tabular}
\end{ruledtabular}
\end{table}

\begin{figure}[tbh]
\centering{
\includegraphics[width=8.5cm]{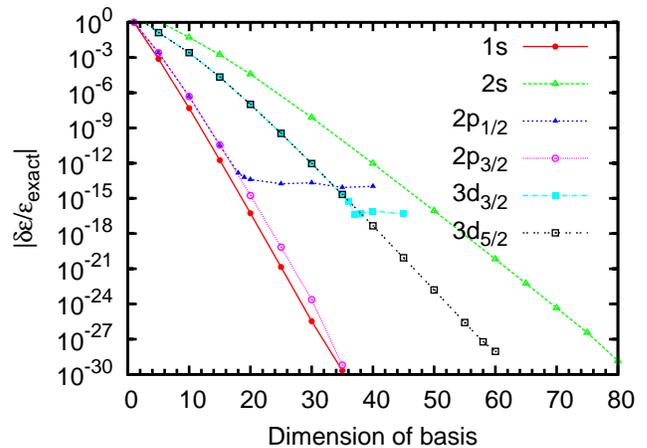}
} \caption{\label{fig:hnpj} (color online). The convergence of the energy
of the low-lying hydrogen eigenstates relative to the exact energy
$\delta_\varepsilon/\varepsilon_\textrm{exact} = (\varepsilon - \varepsilon_\textrm{exact})/\varepsilon_\textrm{exact}$
as the dimension of the L-spinor basis is increased.
The exponent in the L-spinor basis was set to $\lambda = 2.0$.}
\end{figure}

However, we can also compare the basis set convergence of the eigenenergy
to the exact solution of the Dirac equation.  For the states
with $\kappa > 0$ ($2s_{\frac12}$,$2p_{\frac32}$,$3d_{\frac52}$)
the convergence patterns are all monotonic as shown in Fig.~\ref{fig:hnpj}.
Convergence is rapid and an accuracy of about 10$^{-30}$,
is the achievable limit with quadruple precision arithmetic.

The convergence of the eigenenergy for the $2p_{\frac12}$ and
$3d_{\frac32}$ states with increasing dimension of the
L-spinor basis stalled at some point, as also seen in Fig.~\ref{fig:hnpj}.
The $2p_{\frac12}$ energy using the L-spinor representation actually goes
below that of the exact energy at $N = 20$ by $5 \times 10^{-15}$
Hartree.  This is suspicious of a double precision limitation
inside the code for $\kappa < 0$ states.  However, despite
experimentation with both EISPACK and LAPACK eigensolvers
we were unable to push below that of
a purely double precision calculation.  Thus, the remainder of
the Sr$^+$ calculations shown in this paper are all computed
in double precision, where the uncertainties relating to
the core potential lie far above the limits established here.

\subsection{Calculation of transition matrix elements}

The $2^k$-pole oscillator strength, $f_{IJ}^{(k)}$, from
initial state $\Psi_I$ to another eigenstate $\Psi_J$ is defined as
\begin{equation}
f_{IJ}^{(k)}=\frac{2\varepsilon_{IJ}|\langle\Psi_I
\|r^k\textbf{C}^{(k)}(\hat{\textbf{r}})
\|\Psi_J\rangle|^{2}}{(2k+1)(2j_I+1)}
\,,
\label{eqn:oscar}
\end{equation}
with $\varepsilon_{IJ} = E_J - E_I$ being the excitation energy,
$j_I$ is the total angular momentum for the initial state,
and $\textbf{C}^{(k)}(\hat{\textbf{r}})$ is the $k$-th order
spherical tensor.
The line strength,
$|\langle\Psi_I \|r^k\textbf{C}^{(k)}(\hat{\textbf{r}}) \|\Psi_J\rangle|^{2}$,
is calculated via the reduced matrix elements between the orbitals
\begin{equation}
  \langle\Psi_I \|r^k\textbf{C}^{(k)}(\hat{\textbf{r}}) \|\Psi_J\rangle
  =  \sum_{n_I,n_J} c_{n_I} c_{n_J}
  \langle\psi_{n_I} \|r^k\textbf{C}^{(k)}(\hat{\textbf{r}}) \|\psi_{n_J}\rangle ,
\end{equation}
whose (orbital) matrix elements split into a radial part
\begin{equation}
 \begin{split}
 & \langle \psi_{n_I}(r) |r^k | \psi_{n_J}(r)\rangle \\
= & \int_{0}^{\infty}\frac{r^k}{r^{2}}[P_{n_I}(r)P_{n_J}(r)+Q_{n_I}(r)Q_{n_J}(r)]r^2 dr ,
\label{eqn:transradial}
 \end{split}
\end{equation}
multiplied by an angular part~\cite{grant07a}
\begin{eqnarray}
\langle \Omega_{\kappa_I}(\hat{\textbf{r}})
\|\textbf{C}^{(k)}\|\Omega_{\kappa_J }(\hat{\textbf{r}})
\rangle &=&(-1)^{j_I+\frac12} \sqrt{(2j_I+1)(2j_J+1)}
\nonumber \\
&\times& \left (
\begin{array}{ccc}
j_I & j_J & k \\
-\frac12 & \frac12 & 0 \\
\end{array}
\right ).
\label{eqn:transangular}
\end{eqnarray}

\subsubsection{Numerical test: lifetimes of hydrogen atom}

The lifetime of a given state $\Psi_I$
is computed as
\begin{equation}
  \tau_I = \left(\Gamma_I \right)^{-1} = \left(\sum_{k=1}^2 \Gamma^{(k)}\right)^{-1},
\end{equation}
ie. here the decay rate $\Gamma_I$ consists only
of E1 ($k=1$ dipole) and E2 ($k=2$ quadrupole) pathways.
The transition probabilities (in s$^{-1}$) can be written as~\cite{mitroy08c,kelleher08a}
\begin{equation}
A^{(k)} = A_0 \frac{\varepsilon_{IJ}^{2k+1}}{(2j_I+1)c^{2k+1}}
   |\langle\Psi_I \|r^k \textbf{C}^{(k)}(\hat{\textbf{r}})
   \|\Psi_J\rangle|^{2} ,
\label{eqn:transprob1}
\end{equation}
(where the energy differences, the speed of light, and the
matrix elements in this formula are given in atomic units)
The SI unit conversion factor is the inverse of the atomic unit of
time  $A_0 = 1 / (2.418884326509\times 10^{-17}) = 4.1341373336493 \times
10^{16}$ from the latest CODATA\cite{codata15a}.
The results of our calculations are shown in
Table~\ref{tab:hydrogengamma}, where again we also indicate
our agreement with the NIST database~\cite{nistasd500}.
This level of agreement was again expected as we are using
an infinite mass proton approximation when solving the
two-body problem.

\subsection{Calculation of dynamic dipole polarizabilities}

The dynamic dipole ($k=1$) polarizability for a state with angular momentum
$j_I=\frac12$ is independent of the magnetic projection $m_I$,
whilst for $j_I>\frac12$ it depends on $m_I$,
i.e. via scalar ($\alpha_S^{(1)}$) and tensor ($\alpha_T^{(1)}$) components;
\begin{equation}
   \alpha_I^{(1)}(\omega)
 = \alpha_{\mathrm{S}}^{(1)}(\omega)
 + \left(\frac{3m_I^2-j_I(j_I+1)}{j_I(2j_I-1)}\right) \alpha_\mathrm{T}^{(1)}(\omega) .
 \label{eqn:alpham}
\end{equation}

The $2^{k}$-pole scalar polarizability is usually defined in terms
of a sum over all intermediate states, excluding the initial state,
whilst including the continuum~\cite{mitroy10a},
\begin{equation}
\alpha_{\mathrm{S}}^{(k)}(\omega) = \sum_{J \ne I}^{N} \frac{f_{IJ}^{(k)}}{\varepsilon_{IJ}^2-\omega^2} .
\label{eqn:scalaralpha}
\end{equation}
The expression for the tensor part of the dipole polarizability
for a state $I$ can be written as
\begin{eqnarray}
\alpha_{\mathrm{T}}^{(1)}(\omega) &=&
6 \sqrt{\frac{5j_I(2j_I-1)(2j_I+1)}{6(j_I+1)(2j_I+3)}} \nonumber \\
    & \times & \sum_{J \ne I}^{N} (-1)^{j_I+j_J}
  \left\{ \begin{array}{ccc}
   j_I & 1 & j_J \\
   1 & j_J & 2
   \end{array} \right\}
\frac{f^{(1)}_{IJ}}{\varepsilon_{IJ}^2-\omega^2} \,. \quad
\label{eqn:tensoralpha}
\end{eqnarray}
Of interest is mapping out the locations of `tune-out'
wavelengths, $\omega_t$ (where $\alpha_I^{(1)}(\omega_t) \to 0$),
and `magic' wavelengths, $\omega_m$
(where $\left(\alpha_I^{(1)}(\omega_m)-\alpha_J^{(1)}(\omega_m)\right) \to 0$)~\cite{mitroy10a}.

\subsubsection{Numerical test: polarizability of $Z=60$ ion}

A benchmark test of the calculation is to compute the
static dipole polarizability of hydrogenic ion ground states.
The static dipole polarizability of the hydrogenic ground state for $Z = 60$
(excluding negative energy states) is found to be
2.8024692$\times 10^{-7}$ a.u.. This is in agreement to eight significant
digits with a value computed recently using a B-spline basis
\cite{tang12b}.  The same level of agreement is achieved when
negative energy states are included in the polarizability sum rule
\cite{tang12b}.  A similar degree of accuracy is achieved
for the calculation of the quadrupole polarizability.
The quadrupole polarizability of the hydrogenic ground state for $Z = 60$
(including negative energy states) is found to be
2.37114704$\times 10^{-10}$ a.u.. This is in agreement to eight significant
digits with the B-spline value~\cite{tang12b}.

\section{Atomic Properties of S\lowercase{r}$^+$}

Having independently validated the operation of our two codes,
we now turn our attention to the computation of the challenging
one-valence electron ion, Sr$^+$, which requires the
consequent usage of both codes.  First we outline our treatment
of the core-valence interaction.

\subsection{Calculation of the core-valence interaction}

The interaction of the valence electron with the core electrons can be
approximated as a direct and exchange potential, along with a
core-polarization interaction:
\begin{eqnarray}
\hat{V}_{core} \approx \hat{V}_{dir}+\hat{V}_{exc} + \hat{V}_{p_1}.
\end{eqnarray}
A detailed description of the relevant one-body
matrix elements can be found in Ref.~\cite{grant00a}.
In brief, the matrix elements of the direct interaction can be
written as,
\begin{equation}
 \begin{split}
 & \langle \psi_{n_I} |V_{dir} |\psi_{n_J} \rangle \\
 & = \delta_{\kappa_I,\kappa_J}
\int_0^{\infty} \Big( P_{n_I}(r)P_{n_J}(r) + Q_{n_I}(r)Q_{n_J}(r)\Big) V_{d}(r)dr ,
 \end{split}
\end{equation}
where the direct core potential acts locally and radially,
\begin{eqnarray}
V_{d}(r) = \int_0^{r} \frac{\rho_{core}(r^\prime)}{r}dr^\prime +
             \int_{r}^{\infty} \frac{\rho_{core}(r^\prime)}{r^\prime}dr^\prime \, .
\end{eqnarray}
The $\rho_{core}$ is the density of all of the core electrons, where
\begin{eqnarray}
  \rho_{core}(r) = \sum_{c=1}^{N_{core}}(2j_c+1)\left(P_c^2(r)+Q_c^2(r)\right) .
\end{eqnarray}
The $N_{core}$ is the number of core orbitals (denoted by $c$)
obtained from a preceding DF calculation (see Table~\ref{tab:dfenergies}).
The exchange matrix element between the $i$-th and $j$-th valence electron
and the core electrons can be written as a sum over the
interaction with each core electron, viz.
\begin{eqnarray}
\langle \psi_i |V_{exc}
|\psi_{j} \rangle
&= &-\delta_{\kappa_i,\kappa_j}\sum_{c=1}^{N_{core}}\sum_k(2j_c+1)
\nonumber \\
&\times&\left (
\begin{array}{ccc}
j_c  & k & j_i\\
\frac12  & 0 & -\frac12 \\
\end{array}
\right )^2R^k(c,i,j,c) ,
\end{eqnarray}
where
\begin{eqnarray}
R^k(a,b,c,d)=\int_0^{\infty}\int_0^{\infty}(P_a(r_1)P_c(r_1)+Q_a(r_1)Q_c(r_1))  \nonumber \\
    \times\frac{r^k_<}{r^{k+1}_>}(P_b(r_2)P_d(r_2)+Q_b(r_2)Q_d(r_2))dr_1dr_2 . \quad
\end{eqnarray}
Here $r_<$ and $r_>$ are the lesser and greater of the distances $r_1$ and $r_2$ of the
electrons respectively (one of which here is a core electron).  The radial integrals are
computed numerically using Gaussian integration~\cite{bromley01a}, which enables the mixed usage of
Slater-type orbitals (to most compactly represent the core) or Laguerre-type orbitals
(which are orthogonal and thus be included towards completeness without
linear dependence issues).
In order to prevent the valence electrons collapsing into the
core electron (S-spinor only) orbitals, a Gram-Schmidt orthogonalization
of the orbital set is performed to ensure that all the electron
orbitals are orthonormal.

\subsection{Calculation of the semi-empirical potential}

The $e^-$-Sr$^{2+}$ one-body polarization potential $V_{p1}$ is an extension
of the semi-empirical polarization potential used previously~\cite{mitroy08c},
here including dipole, quadrupole, and octupole contributions as
\begin{equation}
V_{p1}(r) = -\sum_{k=1}^{3} \frac{\alpha_{\mathrm{core}}^{(k)}}{2r^{(2(k+1))}}
             \sum_{\ell,j} g^2_{k,\ell,j}(r) | \ell,j \rangle \langle \ell,j| .
\end{equation}
Here, the factors $\alpha_{\mathrm{core}}^{(k)}$ is the static $k$-th order
polarizability of the core electrons (obtained from independent calculations) and
$g^2_{k,\ell,j}(r) = 1- \exp(-r^{(2(k+2))}/\rho^{(2(k+2))}_{\ell,j})$
is a cutoff function designed to make the polarization potential finite at the origin,
while we tune $\rho_{\ell,j}$ for each $\ell,j$ combination.

In our calculations, the core values adopted for the dipole is
$\alpha_{\mathrm{core}}^{(1)} = 5.813$~a.u.~\cite{johnson83a,mitroy03f},
for the quadrupole is $\alpha_{\mathrm{core}}^{(2)} = 17.15$~a.u.~\cite{johnson83a,mitroy03f},
whilst for the octupole is $\alpha_{\mathrm{core}}^{(3)} = 113$~a.u.~\cite{safronova10f}.
The cut-off parameters for the polarization potentials are listed in
Table~\ref{tab:srcutoffs}.  These parameters are set by tuning to the energy of the
lowest state of each $(\ell,j)$ symmetry to the experiment value.
\begin{table}
\caption{\label{tab:srcutoffs} The cutoff parameters, $\rho_{\ell j}$ of the
core polarization potential, for an electron-Sr$^{2+}$ interaction.}
\begin{ruledtabular}
\begin {tabular}{ccccc}
  $\ell$ & $j$ &  $\rho_{\ell j}$ ($a_0$)  & $j$ &  $\rho_{\ell j}$ ($a_0$) \\
\hline
$s$  &  $\frac12$  &   $2.04960$   &  --- &  --- \\
$p$  &  $\frac12$  &   $1.97169$   &  $\frac32$  &   $1.97600$    \\
$d$  &  $\frac32$  &   $2.35353$   &  $\frac52$  &   $2.36534$     \\
$f$  &  $\frac52$  &   $2.15023$   &  $\frac72$  &   $2.19469$     \\
\end{tabular}
\end{ruledtabular}
\end{table}
The dipole transition matrix elements were computed with a modified transition
operator~\cite{hameed68a,hameed72a,mitroy88d,mitroy08c}, e.g.
\begin{equation}
{\bf r} = {\bf r} - \alpha_{core}^{(1)} \sqrt{1 - \exp(-r^6/\bar{\rho}^6)} \frac{\bf r}{r^3}.
\label{dipole}
\end{equation}
The cutoff parameter $\bar{\rho}$ used in Eq.~(\ref{dipole}) was
the average of the $s$, $p$ and $d$ cutoff parameters (note, the
weighting of the $s$ was doubled to give it same weighting as the
two $p$ and $d$ orbitals).

\subsubsection{Results: Energies of Sr$^+$}

For the Sr$^+$ calculations we used Laguerre parameters $\lambda=1.6$ for $s$ orbitals
and $\lambda=1.2$ for the others, with $N=50$ orbitals for each angular momentum.
The energies for a number of low-lying states are given in
Table~\ref{tab:srplusenergies}. Comparing with the experimental data taken from
the National Institute of Science and Technology (NIST)
\cite{nistasd500}, we can find that the error of the present calculations
(labeled as RCICP) is about $2 \times 10^{-4}$~a.u.
for the more highly excited $s$ and $p$ states while being about
five times as large for the $d$ states.
\begin{table}
\caption{\label{tab:srplusenergies} Theoretical (RCICP) and experimental energy levels (in Hartree)
for some of the low-lying states of Sr$^+$. The energies are given relative
to the energy of the Sr$^{2+}$ core. The experimental data are taken from
the National Institute of Science and Technology (NIST) tabulation~\cite{nistasd500}.}
\begin{ruledtabular}
\begin {tabular}{ccccc}
      I  &  $j$   &  $\varepsilon$(RCICP) &  $\varepsilon$(Exp.) &  $\Delta_\varepsilon$ \\
\hline
$5s$ & $\frac12$ &  $-0.4053555$  &  	$-0.4053552$  & $0.0000003$ \\
$4d$ & $\frac32$ &  $-0.3390336$  &  	$-0.3390336$  & $0.0000000$ \\
     & $\frac52$ &  $-0.3377563$  &  	$-0.3377563$  & $0.0000000$ \\
$5p$ & $\frac12$ &  $-0.2973007$  &  	$-0.2973008$  & $0.0000001$ \\
     & $\frac32$ &  $-0.2936464$  &  	$-0.2936491$  & $0.0000027$ \\
$6s$ & $\frac12$ &  $-0.1875380$  &  	$-0.1878515$  & $0.0003135$ \\
$5d$ & $\frac32$ &  $-0.1612581$  &  	$-0.1625649$  & $0.0013068$ \\
     & $\frac52$ &  $-0.1608524$  &  	$-0.1621700$  & $0.0013176$ \\
$6p$ & $\frac12$ &  $-0.1510966$  &  	$-0.1512497$  & $0.0001531$ \\
     & $\frac32$ &  $-0.1497517$  &  	$-0.1499367$  & $0.0001850$ \\
$4f$ & $\frac72$ &  $-0.1274645$  &  	$-0.1274641$  & $0.0000004$ \\
     & $\frac52$ &  $-0.1274582$  &  	$-0.1274582$  & $0.0000000$ \\
$7s$ & $\frac12$ &  $-0.1091774$  &  	$-0.1093570$  & $0.0001796$ \\
$6d$ & $\frac32$ &  $-0.0969695$  &  	$-0.0976983$  & $0.0007288$ \\
     & $\frac52$ &  $-0.0967790$  &  	$-0.0975148$  & $0.0007358$ \\
$7p$ & $\frac12$ &  $-0.0923245$  &     $-0.0924291$  & $0.0001046$ \\
     & $\frac32$ &  $-0.0916778$  &     $-0.0918013$  & $0.0001235$ \\
$5f$ & $\frac52$ &  $-0.0815523$  &  	$-0.0815557$  & $0.0000034$ \\
     & $\frac72$ &  $-0.0815463$  &  	$-0.0815557$  & $0.0000094$ \\
$5g$ & $\frac72$ &  $-0.0802443$  &  	$-0.0802252$  & $0.0000191$ \\
     & $\frac92$ &  $-0.0802442$  &  	$-0.0802252$  & $0.0000190$ \\
\end{tabular}
\end{ruledtabular}
\end{table}

By tuning the polarization potential
cutoff parameters, the spin-orbit splittings are
correct for the $4d_{j}$ and $5p_{j}$ levels.
This also makes reasonably accurate spin-orbit splittings
for the more highly excited states.
Such as, the present calculations
of $6p_{j}$ splitting is $0.001315$~a.u. while the
experimental splitting is $0.001313$~a.u..
The $5d_{j}$ RCICP splitting is $0.000406$~a.u. while the
experimental splitting is $0.000395$~a.u..

\subsection{Line strengths and lifetimes}

The line strengths for a number of low-lying transitions
of Sr$^{+}$ are listed in Table~\ref{tab:srplustrans}.  Line strengths are
mainly given for dipole transitions, while the exceptions are of the
$5s \to 4d_{j}$ transitions.  Table~\ref{tab:srplustrans} also gives
the line strengths from a previous non-relativistic calculation~\cite{mitroy08c},
labeled as CICP, which can be regarded as a precursor to the present
calculation. Finally, Table~\ref{tab:srplustrans} lists the line strengths
of the relativistic all-order single and double many-body
perturbation theory (MBPT-SD) calculation~\cite{jiang09a,safronova10f}.
\begin{table}
\caption{\label{tab:srplustrans}
Comparison of reduced electric dipole (E1) and quadrupole (E2)
line strength for the principal transitions of Sr$^+$ with
other calculations. The $(x)$ notation indicates the error in the last digits.}
\begin{ruledtabular}
\begin {tabular}{lrrr}
Transition              &   RCICP     &  MBPT-SD~\cite{jiang09a,safronova10f}  & CICP~\cite{mitroy08c} \\
\hline
\multicolumn{4}{c}{Dipole} \\
$5s$ - $5p_{\frac12}$       & 9.2852        & 9.474(111)        &  9.2729         \\
$5s$ - $5p_{\frac32}$       & 18.582        & 18.93(22)         &  18.546         \\
$5s$ - $6p_{\frac12}$       & 0.00203       & 0.00063(10)       &  0.000158       \\
$5s$ - $6p_{\frac32}$       & 0.000040      & 0.00116(29)       &  0.000315       \\
$5p_{\frac12}$ - $6s$        & 5.4819        & 5.434(65)         &  5.7963         \\
$5p_{\frac32}$ - $6s$        & 11.903        & 11.81(12)         &  11.593         \\
$6s$ - $6p_{\frac12}$       & 42.681        & 42.64(17)         &  42.414         \\
$6s$ - $6p_{\frac32}$       & 84.392        & 84.29(35)         &  84.827         \\
$6p_{\frac12}$ - $7s$        & 22.763        & 22.77(5)          &  23.964         \\
$6p_{\frac32}$ - $7s$        & 49.132        & 49.07(8)          &  47.928         \\
$5p_{\frac12}$ - $5d_{\frac32}$  & 17.950        & 18.17(32)         &  18.724         \\
$5p_{\frac32}$ - $5d_{\frac32}$  & 3.8161        & 3.869(59)         &  3.7448         \\
$5p_{\frac32}$ - $5d_{\frac52}$  & 33.948        & 34.40(57)         &  33.703         \\
$4d_{\frac32}$ - $5p_{\frac12}$  & 9.5873        & 9.685(181)        &  9.4865         \\
$4d_{\frac32}$ - $5p_{\frac32}$  & 1.9005        & 1.910(36)         &  1.8973         \\
$4d_{\frac52}$ - $5p_{\frac32}$  & 17.409        & 17.53(31)         &  17.076         \\
$4d_{\frac32}$ - $6p_{\frac12}$  & 0.00121       & 0.00608(257)      &  0.00225           \\
$4d_{\frac32}$ - $6p_{\frac32}$  & 0.00111       & 0.00260(76)       &  0.000449          \\
$4d_{\frac52}$ - $6p_{\frac32}$  & 0.00757       & 0.00202(58)       &  0.00404           \\
$4d_{\frac32}$ - $4f_{\frac52}$  & 8.5818        & 8.503(223)        &  8.6472         \\
$4d_{\frac52}$ - $4f_{\frac52}$  & 0.6275        & 0.623(14)         &  0.6177         \\
$4d_{\frac52}$ - $4f_{\frac72}$  & 12.543        & 12.45(30)         &  12.353         \\
\multicolumn{4}{c}{Quadrupole} \\
$5s$ - $4d_{\frac32}$   & 123.04    & 123.94(87)    &  123.08       \\
$5s$ - $4d_{\frac52}$   & 187.50    & 188.98(140)   &  184.63       \\
\end{tabular}
\end{ruledtabular}
\end{table}

The non-relativistic CICP radial matrix elements are the same for the
different members of the same spin-orbit doublets.  So the different
line strength are purely due to geometric factors related to the
angular momentum of the states.   The difference between the CICP
and present RCICP line strengths is typically small, not exceeding $6\%$
for any of the strong transitions. Part of the differences that
occur are due to the different energies of the spin-orbit doublets.
The difference is about $0.1\%$
for the resonance $5s \to 5p_{j}$ transitions.  Differences can be
larger for the weaker transitions with much smaller line strengths
which are much more sensitive to small perturbations in the calculation
of the matrix elements. The generally good agreement
between the CICP and RCICP matrix elements arises because both sets
of calculations have their energies tuned to experimental values.
The binding energy largely determines the long range part of the
wavefunction and it is this part of the wavefunction which dominates
the calculation of the dipole and quadrupole matrix elements.

Our present RCICP calculations generally give
improved results over our previous CICP calculations,
as compared with the MBPT-SD line strengths shown in Table~\ref{tab:srplustrans}.
We now see agreement at the level of a couple of percent between
most of the RCICP and MBPT-SD line strengths,
and most of our results lie within their error estimates.
The RCICP line strengths are $2\%$ smaller than the MBPT-SD line strengths
for the resonant $5s \to 5p_{j}$ transitions, although our results
do lie outside their error estimates~\cite{jiang09a,safronova10f}.
The two most egregious cases are the weak $5s$-$6p_{\frac12}$ and
$4d_{\frac52}$-$6p_{\frac32}$ transitions where we are
around $200\%$ different, even with the relatively large MBPT-SD error
estimates taken into account.  All of the $>2\%$ cases can be
explained again due to the sensitivity to small perturbations
in the calculations.  The Sr$^+$ system presents an extreme
benchmark challenge for all atomic structure methodologies.

Using the line strengths given in Table~\ref{tab:srplustrans},
the lifetimes of $4d_j$ and $5p_j$ states can be easily obtained
using Eqns.~(\ref{eqn:transprob1}).
Table~\ref{tab:srpluslifetimes4d} gives the lifetimes of $4d_j$ states.
The main contribution for the lifetimes of $4d_j$ comes from the
E2 ($4d_j-5s$) transitions. The underlying theoretical framework
of the relativistic coupled cluster (RCC) and MBPT-SD approaches
have many common features~\cite{pal07a,safronova08a,porsev12a}.
In many instances, however, atomic parameters computed using the RCC approach had significant
differences with other independent calculations~\cite{wansbeek08a,wansbeek10a,safronova11a,mitroy08d}.
This situation is also prevalent for the lifetime of the $4d_{j}$ states.
The RCC lifetime ratio $1.1933$ is $8\%$ larger than that given by either the
RCICP and MBPT-SD calculations.  The CICP lifetime ratio of $1.0965$ is
essentially due to the different energies of the two $4d_{j}$ states
(since the matrix elements are the same in the CICP calculation).
The RCICP lifetime ratio $1.1176$ are in excellent agreement with
the MBPT-SD ratio $1.1193$ and most recent experiment ratio $1.115$~\cite{biemont00b}.
\begin{table}[th]
\caption{Lifetimes ($\tau$ in seconds) of the $4d_{\frac32}$ and $4d_{\frac52}$ levels of Sr$^+$
The $4d_{\frac32}:4d_{\frac52}$ lifetime ratio is also given.
}
\label{tab:srpluslifetimes4d}
\begin{ruledtabular}
\begin{tabular}{lccc}
Source                    & $\tau(4d_{\frac32})$  & $\tau(4d_{\frac52})$    & Ratio \\ \hline
RCICP                     & 0.4442       & 0.3974    & 1.1176  \\
RCC~\cite{sahoo06a}       & 0.426(8)     & 0.357(12) & 1.193(65) \\
CICP~\cite{mitroy08c}     & 0.443        & 0.404     & 1.0965   \\
MBPT-SD~\cite{jiang09a}   & 0.441(3)     & 0.394(3)  & 1.119(14) \\
Exp.~\cite{madej90a}      &              & 0.372(25) &    \\
Exp.~\cite{biemont00b}    & 0.455(29)    & 0.408(22) & 1.115(139)   \\
Exp.~\cite{mannervik99a,letchumanan05a} & 0.435(4)   & 0.3908(16) & 1.1131(68) \\
\end{tabular}
\end{ruledtabular}
\end{table}

Different estimates of the $5p_{j}$ lifetimes are given in Table~\ref{tab:srpluslifetimes5p}.
The $5p_{j}$ states have dipole transitions to two lower-lying states,
namely the $5s$ and $4d_{j}$ states. The transition to the
$5s$ state being about twenty times larger than the transition
to the $4d_{j}$ states.
The RCICP and MBPT-SD lifetimes differ by 2$\%$ and the most precise experimental estimates obtained
from laser excitation of ion beams~\cite{kuske78a,pinnington95a}
lie within the RCICP and MBPT-SD estimates. The RCC lifetimes
are smaller than the RCICP and MBPT-SD results. The RCICP and MBPT-SD
comparisons are reminiscent of the $4p_{j}$ lifetimes of Ca$^{+}$.
In Ca$^{+}$ one finds that the RCICP lifetime are about 2$\%$
larger than the MBPT-SD lifetimes~\cite{tang13a}.  The $5p_{\frac12}:5p_{\frac32}$ lifetime
ratio agrees very well with experiment for both calculations.
\begin{table*}[th]
\centering \caption{Lifetimes ($\tau$ in nanoseconds) of the $5p_{\frac12}$ and
$5p_{\frac32}$ states.  The $5p_{\frac12}:5p_{\frac32}$ lifetime ratio is
also given.  The quantity $R$ gives fraction of the total decay rate
arising from the indicated transition.
% \mwjb{has CICP done this before?}
% \jj{No. But we can get it by using the CICP matrix elements.}
}
\label{tab:srpluslifetimes5p}
\begin{ruledtabular}
\begin{tabular}{lccccc}
Level                 & RCICP  &MBPT-SD~\cite{jiang09a} &RCC\cite{kaur15b} & Exp.~\cite{kuske78a} & Exp.~\cite{pinnington95a} \\
 \hline
$\tau(5p_{\frac12})$ & $7.523$ & $7.376$ & $7.16$ & $7.47(7)$ & $7.39(7)$ \\
$R(5p_{\frac12}-5s_{\frac12})$ & $0.9439$ & $0.9444$ & $0.9338$ & &       \\
$R(5p_{\frac12}-4d_{\frac12})$ & $0.0562$ & $0.0556$ & $0.0662$ & &       \\
$\tau(5p_{\frac32})$ & $6.773$ & $6.653$ & $6.44$ & $6.69(7)$ & $6.63(7)$  \\
$R(5p_{\frac32}-5s_{\frac12})$ & $0.9394$ & $0.9400$ & $0.9287$ & &    \\

$R(5p_{\frac32}-4d_{\frac32})$ & $0.0064$ & $0.0064$ & $0.0075$ & &    \\
$R(5p_{\frac32}-4d_{\frac52})$ & $0.0542$ & $0.0536$ & $0.0637$ & &    \\
$5p_{\frac12}:5p_{\frac32}$ Ratio   & $1.111$ & $1.109$ & $1.111$ & $1.117(20)$ & $1.114(20)$ \\
\end{tabular}
\end{ruledtabular}
\end{table*}

\subsection{Static Polarizabilities}

The contributions from the core to the dynamic polarizabilities is only
via a scalar contribution, which was included by a pseudo-oscillator
strength distribution~\cite{margoliash78a,kumar85a,mitroy03f},
\begin{equation}
  \alpha_{\mathrm{S}}^{(core)}(\omega) = \sum_{i}^{N_C} \frac{f_i^{(1)}}{\varepsilon_i^2-\omega^2} .
  \label{eqn:alphacore}
\end{equation}
The pseudo-oscillator strength distribution
is tabulated in Table~\ref{tab:pseudocore}, using the
number of electrons in each shell as the oscillator strength.
Note that in the calculations of polarizability difference for
any two states, the core polarizabilities will effectively cancel each other.
\begin{table}
\caption{Pseudospectral oscillator strength distribution
for the Sr$^{2+}$ core. Energies are given in a.u..
}
\label{tab:pseudocore}

\begin{ruledtabular}
\begin{tabular}{llcc}
$i$ & orbital & $\varepsilon_i$ & $f_i$\\
\hline
1 & $1s^2$    & 583.696    &  2  \\
2 & $2s^2$    &  80.400    &  2   \\
3 & $2p^6$    &  73.005    &  6   \\
4 & $3s^2$    &  13.484    &  2   \\
5 & $3p^6$    &  10.709    &  6   \\
6 & $3d^{10}$ &   5.703    &  10   \\
7 & $4s^2$    &   1.906    &  2   \\
8 & $4p^6$    &   1.108    &  6   \\
\end{tabular}
\end{ruledtabular}
\end{table}

The static dipole and quadrupole polarizabilities of the $5s$, $5p_{j}$ and $4d_{j}$
states are given in Table~\ref{tab:srpluspolar}.
Once again the overall agreement for the dipole polarizability between the RCICP and MBPT-SD calculations is
at the level of $1-2\%$. The present calculations also agree
with the all-order relativistic coupled cluster method with the
singles and doubles approximantion (RCC all-order) results.
The RCICP ground state dipole polarizability of $90.1$~a.u. is about
$2\%$ smaller than the MBPT-SD polarizability. This is the direct consequence of the
slightly different line strengths for the resonant transition.
Since the energies of the lowest eigenstates have been tuned to
the experimental values in both calculations while the slightly
higher difference in the energies of the other excited states has a negligible effect
on the polarizability.
There is only one experimental Sr$^+$ dipole polarizability that has been obtained ~\cite{nunkaew09a}.
In that experiment, the energy differences between the
$5snf$, $5sng$, $5snh$ and $5sni$ states of neutral strontium have been used
to make an estimate of the Sr$^+$ core polarizability. However, the relatively
large uncertainty of $13\%$ cannot be used to discriminate between the different theoretical
estimates.
\begin{table}
\caption{Static ($\omega=0$) scalar and tensor dipole polarizabilities,
$\alpha_{\mathrm{S}}^{(1)}$ and $\alpha_{\mathrm{T}}^{(1)}$,
and static quadrupole polarizabilities $\alpha_{\mathrm{S}}^{(2)}$.
for low-lying states of the Sr$^{+}$ ion.
All numbers are given in a.u..
}
\label{tab:srpluspolar}
\begin{ruledtabular}
\begin{tabular}{lcccc}
   State & term &  RCICP & Others & Others Ref.\\
\hline
$5s_{\frac12}$ & $\alpha_{\mathrm{S}}^{(1)}$ & $90.10$   & $92.2(7)$ & MBPT-SD~\cite{safronova10f} \\
               &                             &           & $91.30$   & MBPT-SD~\cite{jiang09a} \\
               &                             &           & $90.54$   & RCC~all-order~\cite{kaur15a} \\
               &                             &           & $88.29$   & RCC~\cite{sahoo09b} \\
               &                             &           & $89.88$   & CICP~\cite{mitroy08c} \\
               &                             &           & $86(11)$  & Expt.~\cite{nunkaew09a} \\
               & $\alpha_{\mathrm{S}}^{(2)}$ & $1356.27$  & $1370.0(28)$ & MBPT-SD~\cite{safronova10f} \\
               &                             &            & $1346$       & CICP~\cite{mitroy08c} \\
$5p_{\frac12}$ & $\alpha_{\mathrm{S}}^{(1)}$ & $-31.29$  & $-32.2(9)$ & MBPT-SD~\cite{safronova10f} \\
               &                             &           & $-31.27$  & RCC~all-order~\cite{kaur15a} \\
               & $\alpha_{\mathrm{S}}^{(2)}$ & $31595.9$  &              & \\
$5p_{\frac32}$ & $\alpha_{\mathrm{S}}^{(1)}$ & $-20.92$  & $-21.4(8)$ & MBPT-SD~\cite{safronova10f} \\
               &                             &           & $-20.97$  & RCC~all-order~\cite{kaur15a} \\
               & $\alpha_{\mathrm{T}}^{(1)}$ & $9.836$   & $10.74(23)$ & MBPT-SD~\cite{safronova10f} \\
               &                             &           & $10.52$   & RCC~all-order~\cite{kaur15a} \\
               & $\alpha_{\mathrm{S}}^{(2)}$ & $-13098.8$ &              & \\
$4d_{\frac32}$ & $\alpha_{\mathrm{S}}^{(1)}$ & $63.12$   & $63.3(9)$ & MBPT-SD~\cite{safronova10f} \\
               &                             &           & $63.74$  & RCC~all-order~\cite{kaur15a} \\
               &                             &           & $61.43(52)$ & RCC~\cite{sahoo09b} \\
               & $\alpha_{\mathrm{T}}^{(1)}$ & $-35.11$  & $-35.5(6)$   & MBPT-SD~\cite{safronova10f} \\
               &                             &           & $-35.26$     & RCC~all-order~\cite{kaur15a} \\
               &                             &           & $-35.42(25)$ & RCC~\cite{sahoo09b} \\
               & $\alpha_{\mathrm{S}}^{(2)}$ & $2713.8$  &              & \\
$4d_{\frac52}$ & $\alpha_{\mathrm{S}}^{(1)}$ & $61.99$   & $62.0(9)$ & MBPT-SD~\cite{safronova10f} \\
               &                             &           & $62.08$    & RCC~all-order~\cite{kaur15a} \\
               &                             &           & $62.87(75)$ & RCC~\cite{sahoo09b} \\
               & $\alpha_{\mathrm{T}}^{(1)}$ & $-47.38$  & $-47.7(8)$ & MBPT-SD~\cite{safronova10f} \\
               &                             &           & $-47.35$   & RCC~all-order~\cite{kaur15a} \\
               &                             &           & $-48.83(25)$ & RCC~\cite{sahoo09b} \\
               & $\alpha_{\mathrm{S}}^{(2)}$ & $-1728.3$ &            &  \\
\end{tabular}
\end{ruledtabular}
\end{table}

The RCICP quadrupole
polarizability of the ground state is about 1$\%$ smaller than the MBPT-SD polarizability.
The non-relativistic CICP calculation is 2$\%$ smaller than the MBPT-SD polarizability.
This difference is a direct consequence of the difference in the underlying line strengths
between the various calculations.

The RCICP dipole polarizabilities of $5p_j$ agree with MBPT-SD polarizability very well.
The dipole polarizability of $5p_j$ states are negative. That is because the downward
transition from $5p_j$ to the $5s$ and $4d_j$ have very big negative oscillator strengths which
results in the negative polarizability. This is evident in
Table \ref{tab:srplusbreakdown5s5p12} (and Supplemental Tables IV and V),
shows the contributions from different transitions on the polarizabilities.
The tensor dipole polarizability of $5p_{\frac32}$ of RCICP calculations is 8\% samller than the MBPT-SD
calculations. That is mainly because the matrix element of $5s \to 5p_{\frac32}$ of RCICP is samller
than the MBPT-SD matrix element. The RCICP dipole scalar and tensor polariabilities
of $4d_j$ states agree with MBPT-SD and RCC polarizabilities
very well.

One important application of polarizability is to give the magic wavelength by
setting the difference between the polarizabilities of the involved two eigenstates to be zero.
As an example, Table~\ref{tab:srplusstaticpol5s4d} gives the difference of static dipole polarizabilities
for the $5s$ and $4d_j$ states. The polarizability difference between the
$5s$ and $4d_{\frac52}$ is relevant to the determination of the error budget for the
$5s \to 4d_{\frac52}$ clock transition~\cite{madej12a}.  Until recently, the only
estimates of the polarizability difference came from atomic structure
calculations~\cite{mitroy08c,jiang09a,safronova10a,sahoo09b}.
However, the scalar polarizability for this transition has recently
been measured by utilizing the time-dilation effect~\cite{dube14a}.
The time dilation experiment gives a scalar polarizability difference
that lies almost exactly halfway between the RCICP and MBPT-SD
polarizability differences.
\begin{table}
\caption{Difference of static dipole polarizabilities (in a.u.) for the
$5s - 4d_{j}$ transitions of the Sr$^{+}$ ion.}
\label{tab:srplusstaticpol5s4d}
\begin{ruledtabular}
\begin{tabular}{lcc}
Method & $5s-4d_{\frac52}$ & $5s-4d_{\frac32}$  \\
\hline \hline
RCICP &   28.11  &   27.00   \\
MBPT-SD~\cite{safronova10f}  & 30.2  &  28.9   \\
RCC all-order ~\cite{kaur15a}  & 28.46  &  26.8    \\
RCC~\cite{sahoo09b}  & 25.4  &  26.9   \\
Expt.~\cite{dube14a}  & 29.075(43) &     \\
\end{tabular}
\end{ruledtabular}
\end{table}

\subsection{Dynamic polarizabilities and magic wavelengths}

The Sr$^+$ dipole scalar and tensor dynamic polarizabilities are
computed here as per Eqn.~\ref{eqn:alpham}, including the core contribution
as per Eqn.~\ref{eqn:alphacore}. The magic wavelength is calculated by
setting the dynamical polarizability difference between the two involved eigenstates
to be zero.
An example breakdown for the $5s_{\frac12}$ and $5p_{\frac12}$ polarizabilities are
given for both for the static case ($\omega =0$) and also
at the first magic wavelength $\omega = 0.05961933$~a.u.
in Table~\ref{tab:srplusbreakdown5s5p12}.
\begin{table}[th]
\caption{The contributions of individual transitions
to the polarizabilities (in a.u.) of the $5s_{\frac12}$ and $5p_{\frac12}$ states
for the static case and at the magic wavelengths.  These results assume linearly-polarized light.
$\delta \lambda$ are uncertainties
calculated by assuming certain matrix elements have $\pm2\%$ uncertainties.
\label{tab:srplusbreakdown5s5p12}
}
\begin{ruledtabular}
\begin{tabular}{lrr}
$\omega$ (a.u.)   & 0         & 0.05961933 \\
$\lambda$ (nm)   & $\infty$  &  764.2378          \\
$\delta \lambda$ (nm)   &    &  8          \\
Ref.~\cite{kaur15a} (nm)   &    &  769.44          \\
\hline
\multicolumn{3}{c}{$5s_{\frac12}$}            \\
$5p_{\frac12}$      & 28.6439     & 41.1806   \\
$5p_{\frac32}$      & 55.4498     & 77.5362   \\
Remainder       &  0.1891     &  0.1918   \\
Core            &  5.8128     &  5.8276   \\
Total           & 90.0957     &124.7362   \\
\multicolumn{3}{c}{$5p_{\frac12}$ }           \\
$5s_{\frac12}$      &$-$28.6439  &$-$41.1806  \\
$4d_{\frac32}$      &$-$76.5768  &   73.5687  \\
$6s_{\frac12}$      &   16.6954  &   23.7393  \\
$5d_{\frac32}$      &   44.4075  &   55.2192  \\
Remainder       &    7.0082  &   7.5619  \\
Core            &    5.8128  &    5.8276  \\
Total           &$-$31.2969  & 124.7362   \\
\end{tabular}
\end{ruledtabular}
\end{table}

\begin{figure*}[tbh]
\centering{
\includegraphics[width=17cm]{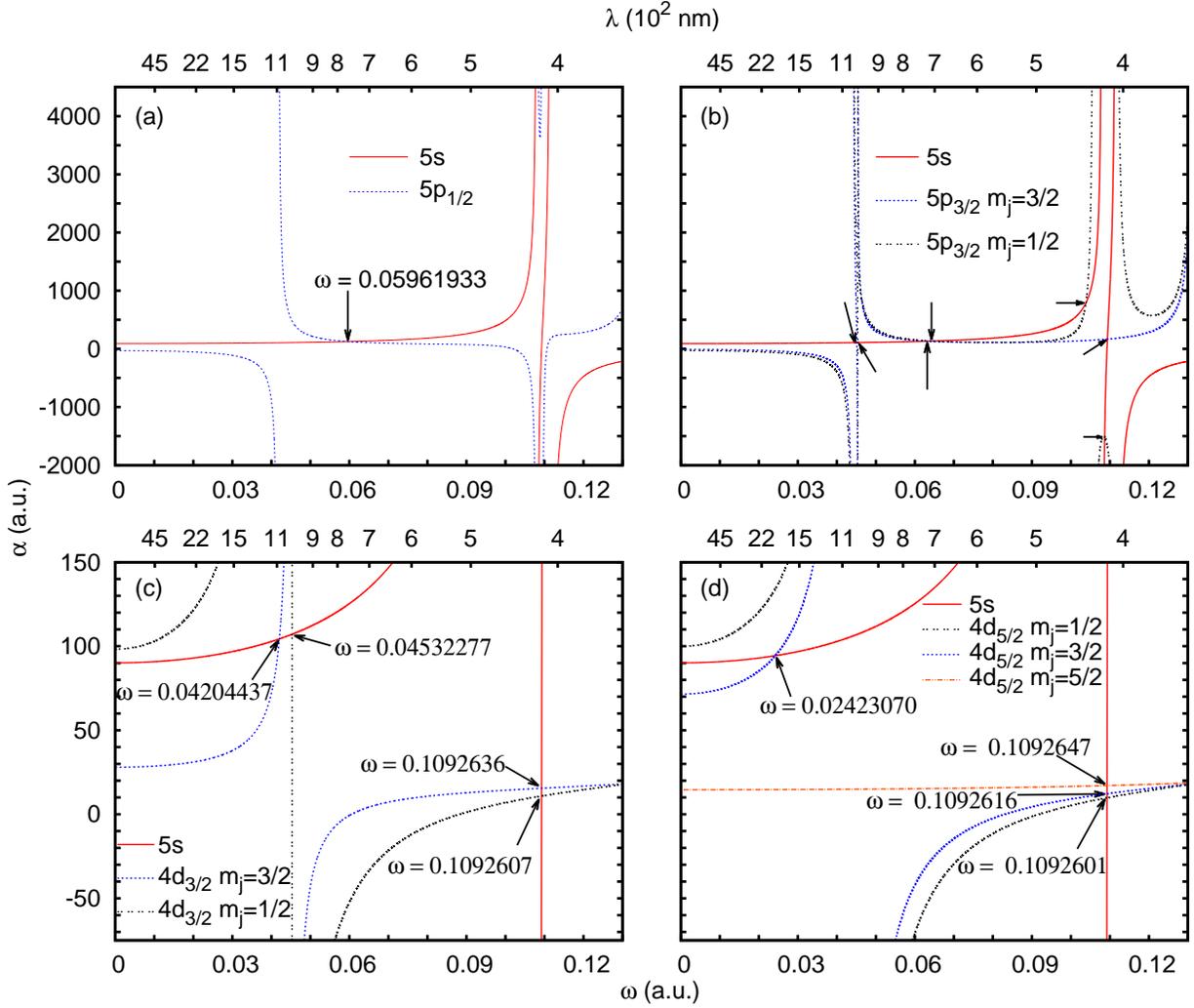}
} \caption{ \label{fig2:multifig} (color online) Dynamic polarizabilities of various states of Sr$^+$.
Panel~(a) compares the $5s_{\frac12}$ and $5p_{\frac12}$ states,
(b) compares $5s_{\frac12}$ and $5p_{\frac32}$,
(c) compares $5s_{\frac12}$ and $4d_{\frac32}$,
(d) compares $5s_{\frac12}$ and $4d_{\frac52}$.
The various magic wavelengths between the respective states are identified by arrows.}
\end{figure*}

%%%%%%%%%%%%%%%%%%%%%%%%%%%%%%%%%%%%%%%%%%%%%%%%%%%%%%%%%%%%%%%%

The dynamic polarizabilities of the $5s_{\frac12}$ and $5p_{\frac12}$ states of Sr$^+$
are shown in Fig.~\ref{fig2:multifig}(a). Note that these calculations assume
linearly-polarized light.  The only magic wavelength for this
transition for wavelengths greater than $400$~nm occurs at $\lambda = 764.238$~nm
($\omega = 0.05961933$~a.u.).
This occurs when the photon energy exceeds the energy for the $5p_{\frac12}$-$4d_{\frac32}$
transition.  While the $5s_{\frac12}$ polarizability is dominated by $5s_{\frac12} - 5p_j$
transition. The breakdown of the $5p_{\frac12}$ polarizability tabulated in
Table~\ref{tab:srplusbreakdown5s5p12} reveals that the transitions to the $5s$, $6s$, $4d_{\frac32}$ and
$5d_{\frac32}$ states all make significant contributions to the $5p_{\frac12}$ polarizability.

The dynamic polarizabilities of the $5s_{\frac12}$ and $5p_{\frac32}$ states of
Sr$^+$ are in Fig.~\ref{fig2:multifig}(b).
Supplemental Tables~IV and V list the breakdown of the polarizabilities
for the static case and at the magic wavelengths for both $m_j$-values.
There are seven magic wavelengths below $\omega = 0.110$~a.u. and
four below 0.070 a.u..
Supplemental Tables~IV and V reveals that the position of the magic
wavelengths near $1004$~nm and $1009$~nm are strongly influenced by the relative sizes of
the $5p_{\frac32} \to 4d_{\frac32}$ and $5p_{\frac32} \to 4d_{\frac52}$ line strengths.
These two magic wavelengths occur when the photon energy lie between the transition energies
of $5p_{\frac32} \to 4d_{\frac32}$ and $5p_{\frac32} \to 4d_{\frac52}$.
Transitions to the $ns_{\frac12}$ states make no contribution to the $5p_{\frac32,m=\frac32}$
state polarizability for linearly polarized light.
Combining with the experimental matrix elements of $5s \to 5p_j$
transitions, the measurement of $1009$~nm magic wavelength
could be able to determine the oscillator strength ratio of
$f_{5p_{\frac32} \to 4d_{\frac32}} : f_{5p_{\frac32} \to 4d_{\frac52}}$. Suppose that all
the remaining components of $5p_{\frac32}$ polarizability
including the $5p_{\frac32} \to 5d_j$ contributions is $5\%$. Then the overall
uncertainty to the polarizability is less than $1\%$.

There are several other magic wavelengths that are worth mentioning.
The magic wavelengths near $709$~nm and $721$~nm
are caused by the gradual increase of
the $5s_{\frac12}$ polarizability as the photon energy approaches the
$5s \to 5p_{j}$ excitation energy and the gradual decrease of the $5p_{\frac32}$
 polarizability as the energy becomes increasingly distant from the
the $5p_{\frac32} \to 4d_{j}$ transition energy.
The magic wavelength at $438$~nm for the $5p_{\frac32,m=\frac12}$ magnetic sub-level
is triggered by the polarizability associated with the
$5p_{\frac32} \to 6s$ transition.
The magic wavelengths near $416$~nm and $419$~nm are caused by the rapid variation of the
$5s$ polarizability for a photon energy lying between the excitation energies from
$5s$ to $5p_{\frac12}$ and $5p_{\frac32}$ states.  These magic wavelengths
can give an estimate of the contribution
to the $5p_{\frac32}$ polarizability arising from excitations to the $nd_{j}$ levels.

The dynamic polarizabilities of the $5s$ and $4d_{\frac52}$ states are shown in
Fig.~\ref{fig2:multifig}(d) while Supplemental Table~VI lists the
breakdown of the polarizabilities for the static case and at the magic wavelengths.
This is probably the most interesting transition since it is the transition of
the Sr$^+$ optical frequency standard.  This
transition has one magic wavelength at $1880$~nm.  This is caused by the increase
in the $4d_{\frac52,m=\frac32}$ polarizability as the photon energy approaches the
$4d_{\frac52} \to 5p_{\frac32}$ excitation energy.
The other three magic wavelengths lie close to $417$~nm and are all caused by
the rapid change of the $5s$ polarizability for photon energies lying
between the excitation thresholds of the $5p_{j}$ doublet.
The magic wavelength mainly arises from the cancellation of the $5p_{\frac12}$
and $5p_{\frac32}$ contributions to the $5s$ dynamic polarizability.
These three magic wavelengths would allow a determination of the oscillator
strength ratio of $f_{5s \to 5p_{\frac12}} : f_{5s \to 5p_{\frac32}}$.
This is similar to Ca$^+$\cite{tang13a,liu15a}, in which
the magic wavelength of $3d_{\frac52} \to 4s_{\frac12}$ clock transition
lying between the transition wavelengths of the $4s \to 4p_j$ doublet was measured and
the ratio of the oscillator strengths $f_{4s \to 4p_{\frac12}}:f_{4s \to 4p_{\frac32}}$
were determined with a deviation of less than 0.5\%.

The dynamic polarizabilities of the $5s$ and $4d_{\frac32}$ states are shown
in Fig.~\ref{fig2:multifig}(c) while  Supplemental Table~VII lists the breakdown of the polarizabilities
for the static case and at the magic wavelengths. This transition has one magic wavelength at $1082$~nm.
This is caused by the increase in the $4d_{\frac32,m=\frac32}$ polarizability
as the photon energy approaches the $4d_{\frac32} \to 5p_{\frac32}$ excitation energy.
Another magic wavelength ($1005$~nm) occurs
at a slightly higher photon energy. It is caused by the rapid change of the
$4d_{\frac32}$ polarizability for photon energies lying between the excitation
energies of the $4d_{\frac32} \to 5p_{j}$ doublet. Combining with the experimental results
of $5s \to 5p_j$ oscillator strength, measurement of this magic wavelength
would give an estimate of the oscillator strength ratio for
$f_{4d_{\frac32} \to 5p_{\frac12}} :f_{4d_{\frac32} \to 5p_{\frac32}}$.

The other two magic wavelengths lie close to $417$~nm are all caused by
the rapid change of the $5s$ polarizability for photon energies lying
between the excitation thresholds of the $5s \to 5p_{j}$ doublet.
Like the magic wavelength near $417$~nm for clock transition $5s \to 5d_{\frac52}$,
measurement of these two magic wavelengths would also
allow a determination of the oscillator
strength ratio for the $5s \to 5p_{\frac12}$ and $5s \to 5p_{\frac32}$ transition.

\subsection{Uncertainties in the magic wavelength positions}

An uncertainty analysis has been done for the magic wavelengths
given in the preceding section.  This analysis estimates how
uncertainties in the matrix elements will translate into changes
in the magic wavelengths.  The motivation for this analysis is to
define reasonable upper and lower limits on the wavelength to assist
an experimental search for these magic wavelengths.

For the $5s \to 5p_j$ polarizability differences, the matrix elements of
$5s \to 5p_j$, $5p_j \to 5s$, $5p_j \to 4d_j$, $5p_j \to 6s$ and
$5p_j \to 5d_j$ are dominant. For the $5s \to 4d_j$ polarizability differences,
the $5s \to 5p_j$ and $4d_j \to 5p_j$ matrix elements are dominant.
All these matrix elements were changed by 2\% (as most of the reliable calculations
and experiments agree with each other within a 2\% difference) and the magic
wavelengths were recomputed.  The resultant difference is set as the uncertainty
of the magic wavelength.  The matrix elements involving the different spin-orbit
states of the same multiplet were all given the same scaling.

The uncertainties of each magic wavelength is given
in Table~\ref{tab:srplusbreakdown5s5p12}
(and Supplemental Tables IV, V, VI and VII).
It can be found that the magic wavelengths
$764$~nm for $5s \to 5p_{\frac12}$,
$709$~nm for $5s \to 5p_{\frac32 m=\frac12}$,
$721$~nm for $5s \to 5p_{\frac32 m=\frac32}$,
$1083$~nm for $5s \to 4d_{\frac32 m=\frac32}$, and
$1880$~nm for $5s \to 4d_{\frac52 m=\frac32}$ are
relatively sensitive to change in the matrix element. The uncertainties of
magic wavelength are from $4$~nm to $133$~nm.  The reason is that the rate of
change of $5s$ and $5p_j$ (or $4d_j$) polarizabilities are small
near these magic wavelength, namely $d\alpha/d\omega$ are small.
The magic wavelength calculated by Kaur \textit{et al.}~\cite{kaur15a} using RCC all-order method lie in our uncertainties.

There are some of the magic wavelengths, however, such as
$1009$~nm, $1004$~nm, and $417$~nm, that are relatively insensitive
to the changes of matrix elements.  The magic wavelength $1009$~nm, $1004$~nm lie in
between the transition energy of $5p_{\frac32} \to 4d_j$ spin-orbit doublet.
The magic wave lengths near the $417$~nm lie between the transition energy of
$5s \to 5p_j$ spin-orbit doublet.  Present calculations of magic wavelengths agree with
RCC all-order results of Kaur \textit{et al.}~\cite{kaur15a} excellently.

Experimental determination of the oscillator strengths for the resonant
$5s \to 5p_{j}$ transitions using a lifetime approach is complicated due to
the existence of the $4d_{j} \to 5p_{j}$ transitions.  However, the
measurement of magic wavelengths near $417$~nm for the $5s \to 4d_{j}$ transitions
can give a reasonable estimate of the oscillator ratio of the two
transitions of the $5s \to 5p_{j}$ doublet since the polarizability
of the $5s$ state is so much larger than any of the other polarizabilities.
The magic wavelength at $416.9999$~nm changes by $0.0002$~nm when the
$4d_{\frac52}-4f_j$ matrix elements are changed by $2\%$.

\section{Conclusion}

The development and realization of a relativistic model potential description
of quasi single electron atoms and ions was presented.  Rather than using a
B-spline basis~\cite{johnson88a}, the single electron spinors are expanded as a linear
combination of S-spinors and L-spinors. The starting point of the calculation
is a DF calculation for the core state. The DF wavefunctions then serve as
a starting point for the calculations to describe the ground and excited states
of quasi single electron atoms or ions. The core electrons are kept frozen,
where the direct and exchange interactions between the valence electron
and the core are computed without approximation. Dynamical interactions between the
valence electron and the core beyond the DF level are incorporated through
semi-empirical polarization potentials.

The method was applied to
the description of the low-lying states of Sr$^+$ giving line strengths
and polarizabilities that are generally within $1-2\%$ of the significantly
more computationally demanding relativistic all-order singles and doubles
method~\cite{safronova10f,jiang09a}.
A number of magic wavelengths are identified for the
$5s \to 5p_{\frac12}$, $5s \to 5p_{\frac32}$, $5s \to 4d_{\frac32}$ and $5s \to 4d_{\frac52}$
transitions. We recommend that the measurements of $1009$~nm magic wavelength
for the $5s\to 5p_{\frac32}$ transition could be able to determine the
oscillator strength ratio of $f_{5p_{\frac32} \to 4d_{\frac32}} : f_{5p_{\frac32} \to 4d_{\frac52}}$.
The determination of the magic wavelengths for the
$5s \to 4d_{\frac32}$ and $5s \to 4d_{\frac52}$ transitions near $417$~nm
 would allow a determination of the oscillator
strength ratio for the $5s \to 5p_{\frac12}$ and $5s \to 5p_{\frac32}$ transitions.

This approach can also be used for a variety of heavy atoms or ions,
such as Cs, Ba$^+$, Yb$^+$, and so on.
Atomic properties including the energy levels, the oscillator
strengths, the static and dynamic multipole polarizabilities,
the black-body radiation shifts, and the dispersion coefficients that
characterize the long-range interaction between pairs of atoms,
can be studied with improved accuracy over our previous
non-relativistic-based CICP treatment~\cite{mitroy03f}.

\begin{acknowledgments}
This research was partially supported by the Australian Research Council (ARC)
Discovery Project DP1092620. The work of JJ was supported by
National Natural Science Foundation of China (NSFC) (Grants No.11147018 and 11564036)
and YC was supported by NSFC (Grants No.11304063). The work of MWJB was
supported by an ARC Future Fellowship (FT100100905).
The authors would like to thank Prof. Ulyana Safronova for providing tables
of reduced matrix elements for Sr$^{+}$.
We would like to thank Prof. Fumihiro Koike for his valuable suggestions.
\end{acknowledgments}

%%%%%%%%%%%%%%%%%%%%%%%% begin thebibliography  %%%%%%%%%%%%%%%%%%%%%%%%%%

%\bibliography{amo-darwin}

\newpage

\appendix

\subsection{Supplemental: Dirac-Fock basis sets}
\label{app:dfbasis}

In Tables~\ref{tab:dfsspinors1}, \ref{tab:dfsspinors2}, and \ref{tab:dfsspinors3}
we list the optimized S-spinor basis sets that were used to
compute the energies shown in the main paper Table~I.
\begin{table*}[th]
\caption{The exponents, $\lambda_i$, of S-spinor basis for the closed-shell state of the alkali-metal atoms}
\label{tab:dfsspinors1}
\begin{ruledtabular}
\begin{tabular}{rrrrrrrrrrrr}
Li$^+$     &   Na$^+$      &             & K$^+$           &                &   Rb$^+$     &             &            & Cs$^+$     &            &             \\
7$s$       &   7$s$        &   4$p$      & 8$s$            &  7$p$          &   11$s$      &   8$p$      &   5$d$     & 14$s$      & 12$p$      &  10$d$      \\
15.00000   &  17.478901    &   8.906343  &   22.790000     &   21.700000    &   54.542112  &  36.000000  & 20.676607  &117.539065  &  50.898116 &  44.421251  \\
8.00000    &  10.588106    &   4.706615  &   17.292481     &   11.552057    &   36.714627  &  28.566009  & 11.413245  & 83.361039  &  36.617350 &  31.729465  \\
4.69873    &   7.376552    &   2.516997  &   10.488771     &    7.027067    &   30.744640  &  17.212806  &  7.057890  & 59.121304  &  26.343417 &  22.663903  \\
2.47673    &   3.908322    &   1.581117  &    5.372695     &    5.273478    &   17.639600  &  12.762750  &  4.679490  & 41.930003  &  18.952099 &  16.188502  \\
1.63200    &   2.644974    &             &    4.612349     &    3.189831    &   14.845074  &   5.836649  &  3.013206  & 29.737591  &  13.634603 &  11.563216  \\
1.07000    &   2.016045    &             &    2.685122     &    2.301755    &    7.435420  &   4.475461  &            & 21.090490  &  9.809067  &   8.259440  \\
0.66055    &   1.084267    &             &    1.797145     &    1.641334    &    6.200901  &   1.892352  &            & 14.957794  &  7.056883  &   5.899600  \\
           &               &             &    0.969078     &                &    3.547780  &   1.413277  &            & 10.608365  &  5.076894  &   4.214000  \\
           &               &             &                 &                &    2.285862  &             &            &  7.523663  &  3.652442  &   3.010000  \\
           &               &             &                 &                &    1.729660  &             &            &  5.335931  &  2.627656  &   2.150000  \\
           &               &             &                 &                &    1.462792  &             &            &  3.784348  &  1.890400  &             \\
           &               &             &                 &                &              &             &            &  2.683935  &  1.360000  &             \\
           &               &             &                 &                &              &             &            &  1.903500  &            &             \\
           &               &             &                 &                &              &             &            &  1.350000  &            &             \\
\end{tabular}
\end{ruledtabular}
\end{table*}
\begin{table*}[th]
\caption{The exponents, $\lambda_i$, of S-spinor basis for the closed-shell state of the nobel gas atoms}
\label{tab:dfsspinors2}
\begin{ruledtabular}
\begin{tabular}{rrrrrrrrrrr}
 Ne        &               &      Ar     &                 &    Kr          &              &             & Xe         &            &         \\
   8$s$    &   5$p$        &   11$s$     &    9$p$         &    10$s$       &   9$p$       &5$d$         & 14$s$      & 12$p$      &  9$d$   \\
 45.399135 & 8.478657      & 34.980000   &    21.819460    &   55.509710    &   29.485877  &   18.970734 & 75.251507 & 35.791162 & 32.432647 \\
 26.975125 & 4.900173      & 27.650027   &    16.993680    &   35.893641    &   17.401415  &   10.615817 & 53.588918 & 25.360241 & 22.627481 \\
 16.028001 & 2.832016      & 19.938400   &    10.964770    &   30.521126    &   14.896787  &    6.554858 & 38.160646 & 17.969292 & 15.786651 \\
 9.523470  & 1.636742      & 16.931250   &     7.482350    &   16.222298    &    9.460614  &    4.324848 & 27.174179 & 12.732349 & 11.013968 \\
 5.658628  & 0.945942      & 12.295700   &     6.139970    &   15.137969    &    6.280458  &    2.647825 & 19.350721 &  9.021653 &  7.684182 \\
 3.362227  &               &  7.384180   &     3.380570    &    9.419056    &    4.517666  &             & 13.779640 &  6.392397 &  5.361070 \\
 1.997758  &               &  5.871280   &     2.775260    &    6.515549    &    3.649240  &             &  9.812476 &  4.529406 &  3.740290 \\
 1.187022  &               &  3.985270   &     1.471470    &    3.780907    &    2.269718  &             &  6.987460 &  3.209363 &  2.609511 \\
           &               &  1.519523   &     0.994720    &    2.580124    &    1.453805  &             &  4.975767 &  2.274031 &  1.820593 \\
           &               &  1.290069   &                 &    1.826235    &              &             &  3.543242 &  1.611291 &           \\
           &               &  0.897910   &                 &                &              &             &  2.523141 &  1.141699 &           \\
           &               &             &                 &                &              &             &  1.796728 &  0.808964 &           \\
           &               &             &                 &                &              &             &  1.279449 &           &           \\
           &               &             &                 &                &              &             &  0.911095 &           &           \\
\end{tabular}
\end{ruledtabular}
\end{table*}
\begin{table*}[th]
\caption{The exponents, $\lambda_i$, of S-spinor basis for the closed-shell state of the alkaline-earth atoms}
\label{tab:dfsspinors3}
\begin{ruledtabular}
\begin{tabular}{rrrrrrrrrrr}
 Be$^{2+}$ &  Mg$^{2+}$ &             &  Ca$^{2+}$    &                 &  Sr$^{2+}$     &           &            &   Ba$^{2+}$    &           &      \\
 5$s$      &   9$s$     &   6$p$      & 10$s$         &    9$p$         &    12$s$       &    10$p$  &     5$d$   &    15$s$       &    14$p$  &   10$d$  \\
 20.645400 & 20.000000  & 20.291000   &   39.220281   & 59.809200       &  78.966521     & 35.880469 & 21.271807  & 116.958344     & 47.985577 &  45.454303    \\
  6.548210 & 12.980000  &  9.086105   &   27.331206   & 18.309400       &  56.404658     & 23.920312 & 11.766205  &  85.060614     & 35.544872 &  32.467359    \\
  3.521360 & 10.987200  &  5.384429   &   19.046137   & 11.621200       &  40.289041     & 15.946875 &  7.246050  &  61.862265     & 26.329535 &  23.190971    \\
  2.423980 &  6.826120  &  3.016046   &   13.272569   &  7.243230       &  28.777887     & 10.631250 &  4.827920  &  44.990738     & 19.503359 &  16.564979    \\
  0.733843 &  2.808800  &  1.485000   &    9.249177   &  5.090000       &  20.555633     &  7.087500 &  3.105406  &  32.720537     & 14.446933 &  11.832128    \\
           &  1.982660  &  1.352222   &    6.445420   &  4.430000       &  14.682595     &  4.725000 &            &  23.796754     & 10.701432 &   8.451520    \\
           &  1.209417  &             &    4.491582   &  2.677958       &  10.487568     &  3.150000 &            &  17.306730     &  7.926986 &   6.036800    \\
           &  0.986366  &             &    3.130022   &  2.063877       &   7.491120     &  2.100000 &            &  12.586713     &  5.871842 &   4.312000    \\
           &  0.751237  &             &    2.181200   &  1.489720       &   5.350800     &  1.400000 &            &   9.153973     &  4.349512 &   3.080000    \\
           &            &             &    1.520000   &                 &   3.822000     &  2.600000 &            &   6.657435     &  3.221861 &   2.200000    \\
           &            &             &               &                 &   2.730000     &           &            &   4.841771     &  2.386564 &               \\
           &            &             &               &                 &   1.950000     &           &            &   3.521288     &  1.767825 &               \\
           &            &             &               &                 &                &           &            &   2.560937     &  1.309500 &                \\
           &            &             &               &                 &                &           &            &   1.862499     &  0.970000 &                \\
           &            &             &               &                 &                &           &            &   1.354545     &           &     \\
\end{tabular}
\end{ruledtabular}
\end{table*}

\subsection{Supplemental: polarizability breakdowns}
\label{app:polbreakdowns}

We present in Table~\ref{tab:srplusbreakdown5s5p32m12} the breakdowns of the
$5s_{\frac12}$ and $5p_{\frac32}$ (for $m_j = \frac12$) at both $\omega=0$ and at the
magic wavelengths.
\begin{table*}[th]
\caption{The contributions of individual transitions
to the polarizabilities (in a.u.) of the $5s_{\frac12}$ and $5p_{\frac32,m=\frac12}$ states
at the magic wavelengths.  These results assume linearly-polarized light.
$\delta \lambda$ are uncertainties
calculated by assuming certain matrix elements have $\pm2\%$ uncertainties.
}
\label{tab:srplusbreakdown5s5p32m12}
\begin{ruledtabular}
\begin{tabular}{lrrrrr}
$\omega$ (a.u.)  & 0         & 0.04536049 & 0.06425984    &  0.1038352  & 0.1086593 \\
$\lambda$ (nm)   & $\infty$  & 1004.4722      & 709.0487         & 438.8044    &419.3232  \\
$\delta \lambda$ (nm)   &    &    0.02        & 15     & 0.5   &   0.02    \\
Ref.~\cite{kaur15a} (nm)   &    &1004.47     & 716.72     &438.37  & 419.30    \\
      \hline
\multicolumn{5}{c}{$5s_{\frac12}$}    \\
$5p_{\frac12}$      & 28.6439    & 34.7716    &    44.3176    &  374.0870  &$-$2551.0885  \\
$5p_{\frac32}$      & 55.4498    & 66.3984    &    82.8749    &  407.8492  &   1030.5370  \\
$6p_{\frac12}$      &  0.0027    &  0.0027    &     0.0028    &    0.0032  &      0.0033  \\
$6p_{\frac32}$      &  0.0001    &  0.0001    &     0.0001    &    0.0001  &      0.0001  \\
Remainder       &  0.1864    &  0.1879    &     0.1894    &    0.1945  &      0.1953  \\
Core            &  5.8128    &  5.8214    &     5.8300    &    5.8579  &      5.8623  \\
Total           & 90.0957    &107.1821    &   133.2148    &  788.0019  &$-$1514.4906  \\
\multicolumn{5}{c}{$5p_{\frac32}$ $(m=\frac12)$}    \\
$5s_{\frac12}$      &$-$55.4498    &  $-$66.3984   & $-$82.8749    &$-$407.8492   &$-$1030.5370 \\
$4d_{\frac32}$      & $-$1.3958    &$-$1317.0076   &     1.3892    &     0.3296   &      0.2950 \\
$4d_{\frac52}$      &$-$78.9377    &   1369.5704   &    70.3193    &    17.3793   &     15.5727 \\
$6s_{\frac12}$      &   37.5051    &     45.9523   &    59.4297    &  1020.4421   & $-$684.0269 \\
$5d_{\frac32}$      &    0.9704    &      1.1025   &     1.2774    &     2.6050   &      3.1017 \\
$5d_{\frac52}$      &   51.6420    &     58.6193   &    67.8494    &   137.2366   &    162.9085 \\
Remainder       &    9.0997    &      9.5223   &     9.9947    &    11.9506   &     12.3331 \\
Core            &    5.8128    &      5.8214   &     5.8300    &     5.8579   &      5.8623 \\
Total           &$-$30.7531    &    107.1821   &   133.2148    &   788.0019   &$-$1514.4906 \\
\end{tabular}
\end{ruledtabular}
\end{table*}

We present in Table~\ref{tab:srplusbreakdown5s5p32m32} and breakdowns of the
$5s_{\frac12}$ and $5p_{\frac32}$ (for $m_j = \frac32$) at both $\omega=0$ and at the
magic wavelengths.
\begin{table*}[th]
\caption{The contributions of individual transitions
to the polarizabilities (in a.u.) of the $5s_{\frac12}$ and $5p_{\frac32,m=\frac32}$ states
at the magic wavelengths.  These results assume linearly-polarized light.
$\delta \lambda$ are uncertainties
calculated by assuming certain matrix elements have $\pm2\%$ uncertainties.
}
\label{tab:srplusbreakdown5s5p32m32}
\begin{ruledtabular}
\begin{tabular}{lrrrr}
$\omega$ (a.u.) & 0          & 0.04512102   & 0.06316610      &  0.1093628 \\
$\lambda$ (nm)  & $\infty$   & 1009.8032     & 721.3260         & 416.6258  \\
$\delta \lambda$ (nm)  &     &  0.2            &  10 & 0.02      \\
Ref.~\cite{kaur15a} (nm)  &    & 1009.80  & 724.92   & 416.62  \\
\hline
\multicolumn{5}{c}{$5s_{\frac12}$}    \\
$5p_{\frac12}$      & 28.6439     & 34.6935    &    43.5140    &$-$1175.6343  \\
$5p_{\frac32}$      & 55.4498     & 66.2606    &    81.5142    &   1335.6567 \\
$6p_{\frac12}$      &  0.0027     &  0.0027    &     0.0028    &      0.0033 \\
$6p_{\frac32}$      &  0.0001     &  0.0001    &     0.0001    &      0.0001 \\
Remainder       &  0.1864     &  0.1878    &     0.1893    &      0.1954 \\
Core            &  5.8128     &  5.8213    &     5.8294    &      5.8629 \\
Total           & 90.0957     &106.9660    &   131.0497    &    166.0840 \\
\multicolumn{5}{c}{$5p_{\frac32}$ $(m=\frac32)$}    \\
$5s_{\frac12}$      &    0.0000  &      0.0000  &      0.0000    &   0.0000   \\
$4d_{\frac32}$      &$-$12.5624  &$-$1084.8929  &     13.4056    &   2.6136   \\
$4d_{\frac52}$      &$-$52.6251  &   1131.7423  &     50.0752    &  10.2228 \\
$6s_{\frac12}$      &    0.0000  &      0.0000  &      0.0000    &   0.0000 \\
$5d_{\frac32}$      &    8.7340  &      9.9079  &     11.3754    &  28.7347 \\
$5d_{\frac52}$      &   34.4280  &     39.0240  &     44.7588    & 111.7330   \\
Remainder       &    5.1321  &      5.3621  &      5.6053    &   6.9170 \\
Core            &    5.8128  &      5.8213  &      5.8294    &   5.8629 \\
Total           &$-$11.0806  &    106.9660  &    131.0497    & 166.0840   \\
\end{tabular}
\end{ruledtabular}
\end{table*}

We present in Table~\ref{tab:srplusbreakdown5s4d52} the breakdowns of the
$5s_{\frac12}$ and $4d_{\frac52}$ at both $\omega=0$ and at the
magic wavelengths.
\begin{table*}[H]
\centering \caption{ The contributions of individual transitions
to the polarizabilities (in a.u.) of the $5s_{\frac12}$ and $4d_{\frac52}$ states
at the magic wavelengths.  These results assume linearly-polarized light.
$\delta \lambda$ are uncertainties
calculated by assuming certain matrix elements have $\pm2\%$ uncertainties.
}
\label{tab:srplusbreakdown5s4d52}
\begin{ruledtabular}
\begin{tabular}{lrrrrr}
$\omega$ (a.u.)    &$0$          & 0.02423070 & 0.1092601 & 0.1092616  & 0.1092647   \\
$\lambda$(nm)      &$\infty$     & 1880.3976   &   417.0175  &  417.0116    &   416.9999   \\
$\delta \lambda$ (nm)  &     &  103             &  0.0014 & 0.0016 & 0.0010    \\
Ref.~\cite{kaur15a}(nm) &$\infty$ &              &   417.01  &  417.00    &   417.00   \\
\hline
  &  \multicolumn{5}{c}{$5s_{\frac12}$}\\
$5p_{\frac12}$         & 28.6439      &30.1606     &$-$1276.3995     &$-$1274.7701    &$-$1271.5180    \\
$5p_{\frac32}$         & 55.4498      &58.1876     &   1280.1591     &   1280.9529    &   1282.5461    \\
Remainder            &  0.1891      & 0.1895     &      0.1988     &      0.1988    &      0.1988    \\
Core                 &  5.8128      & 5.8153     &      5.8628     &      5.8628    &      5.8628    \\
Total                & 90.0957      &94.3530     &      9.8211     &     12.2444    &     17.0897    \\
 &    \multicolumn{5}{c}{$4d_{\frac52}$}\\
 &                 Average  & $m_{j}=\frac32$ & $m_{j}=\frac12$    & $m_{j}=\frac32$      & $m_{j}=\frac52$    \\
$5p_{\frac32}$         & 43.8543    &75.3721     &$-$15.3687     &$-$10.2455       &  0.0000       \\
$6p_{\frac32}$         &  0.0045    & 0.0055     &    0.0122     &    0.0081       &  0.0000       \\
$4f_{\frac52}$         &  0.3316    & 0.2592     &    0.0389     &    0.3503       &  0.9732         \\
$4f_{\frac72}$         &  6.6273    & 7.1963     &   11.6713     &    9.7262       &  5.8358         \\
$7p_{\frac32}$         &  0.0009    & 0.0010     &    0.0019     &    0.0013       &  0.0000         \\
$5f_{\frac52}$         &  0.0859    & 0.0669     &    0.0090     &    0.0810       &  0.2250         \\
$5f_{\frac72}$         &  1.7167    & 1.8560     &    2.6979     &    2.2483       &  1.3490         \\
Remainder            &  3.5576    & 3.7808     &    4.8957     &    4.2118       &  2.8439         \\
Core                 &  5.8128    & 5.8153     &    5.8628     &    5.8628       &  5.8628         \\
Total                & 61.9916    &94.3530     &    9.8211     &   12.2444       & 17.0897        \\
\end{tabular}
\end{ruledtabular}
\end{table*}

We present in Table~\ref{tab:srplusbreakdown5s4d32} the breakdowns of the
$5s_{\frac12}$ and $4d_{\frac32}$ at both $\omega=0$ and at the
magic wavelengths.
\begin{table*}[H]
\caption{ The contributions of individual transitions
to the polarizabilities (in a.u.) of the $5s_{\frac12}$ and $4d_{\frac32}$ states
at the magic wavelengths.  These results assume linearly-polarized light.
$\delta \lambda$ are uncertainties
calculated by assuming certain matrix elements have $\pm2\%$ uncertainties.
\label{tab:srplusbreakdown5s4d32}
}
\begin{ruledtabular}
\begin{tabular}{lrrrrr}
 $\omega$(a.u.)           &0             & 0.04204437  & 0.04532277  & 0.1092607  & 0.1092636 \\
 $\lambda$(nm)             &$\infty$      & 1083.6969   & 1005.3081    & 417.0149     & 417.0038      \\
$\delta \lambda$ (nm)      &             &  4            &  0.011          & 0.0017    & 0.0012      \\
Ref.~\cite{kaur15a} (nm)     &              & 1082.38    & 1005.30    & 417.00     & 417.00      \\
\hline
  & \multicolumn{5}{c}{$5s_{\frac12}$} \\
${5p_{\frac12}}$         & 28.6439     & 33.7544    &  34.7592   &$-$1275.7000 &$-$1272.5909 \\
${5p_{\frac32}}$         & 55.4498     & 64.6016    &  66.3766   &   1280.4995 &   1282.0191 \\
Remainder            &  0.1891     &  0.1904    &   0.1907   &      0.1988 &      0.1988 \\
core                 &  5.8128     &  5.8202    &   5.8214   &      5.8628 &      5.8628 \\
Total                & 90.0957     &104.3666    & 107.1479   &     10.8611 &     15.4898 \\
  &  \multicolumn{5}{c}{$4d_{\frac32}$}\\
                     &Average     &$m_{j}=\frac32$ &$m_{j}=\frac12$  &$m_{j}=\frac12$  &$m_{j}=\frac32$ \\
${5p_{\frac12}}$         &  38.2884      &   0.0000     &$-$426.7411  &$-$13.0801 &   0.0000  \\
${5p_{\frac32}}$         &   6.9791      &  88.6061     &   513.0793  & $-$0.2911 &$-$2.6193  \\
${6p_{\frac12}}$         &   0.0011      &   0.0000     &     0.0023  &    0.0032 &   0.0000   \\
${6p_{\frac32}}$         &   0.0010      &   0.0018     &     0.0002  &    0.0003 &   0.0026 \\
${4f_{\frac52}}$         &   6.7604      &   5.6307     &     8.5027  &   11.0630 &   7.3755 \\
${7p_{\frac12}}$         &   0.0002      &   0.0000     &     0.0004  &    0.0005 &   0.0000 \\
${7p_{\frac32}}$         &   0.0002      &   0.0003     &     0.0000  &    0.0000 &   0.0004 \\
${5f_{\frac52}}$         &   1.7645      &   1.4503     &     2.1852  &    2.5826 &   1.7217 \\
Remainder            &   3.5164      &   2.8571     &     4.2974  &    4.7197 &   3.1460 \\
core                 &   5.8128      &   5.8202     &     5.8214  &    5.8628 &   5.8628 \\
Total                &  63.1242      & 104.3666     &   107.1479  &   10.8611 &  15.4898 \\
\end{tabular}
\end{ruledtabular}
\end{table*}

\end{document}